\documentclass[runningheads]{llncs}

\usepackage{eccv}

\usepackage{eccvabbrv}

\usepackage{graphicx}
\usepackage{booktabs}

\usepackage[accsupp]{axessibility}  

\usepackage{adjustbox}
\usepackage{multirow}
\usepackage{array}

\usepackage{algorithm}
\usepackage{algcompatible}
\usepackage{tabularx}
\algnewcommand\algorithmicreturn{\textbf{return}}
\algnewcommand\RETURN{\State \algorithmicreturn}

\usepackage{hyperref}

\usepackage{orcidlink}

\begin{document}

\title{Region-Aware Multimodal Large Language Model via SlowFast Tokenization and Pseudo-Mask Guidance for 3D CT Report Generation} 

\author{Sunggu Kyung\inst{1}\orcidlink{0000-0002-7582-9484} \and
Jinyoung Seo\inst{2} \and
Hyunseok Lim\inst{1} \and
Dongyeong Kim\inst{3} \and
Hyungbin Park\inst{2} \and
Jimin Sung\inst{2} \and
Wooyoung Jo\inst{2} \and
Yoojin Nam\inst{3} \and
Namkug Kim\inst{1}\thanks{Corresponding author}}

\titlerunning{MedRegion-CT for 3D CT Report Generation}
\authorrunning{Kyung et al.}

\institute{
Department of Convergence Medicine, University of Ulsan College of Medicine, Asan Medical Center, Seoul, Republic of Korea \and
University of Ulsan College of Medicine, Seoul, Republic of Korea \and
Department of Radiology and Research Institute of Radiology, University of Ulsan College of Medicine, Asan Medical Center, Seoul, Republic of Korea \\
\email{babbu3682@gmail.com, namkugkim@gmail.com}}

\maketitle

\begin{abstract}
  Current CT report generation frameworks predominantly rely on global feature representations, often failing to capture region-specific details and potentially missing certain abnormalities. To overcome this limitation, we propose MedRegion-CT, a region-focused multimodal large language model framework featuring three key innovations. First, we revisit the SlowFast strategy to jointly model global and fine-grained information and adapt it to the medical domain via a Region-based SlowFast Tokenizer that extracts tokens guided by clinically meaningful regions. Second, generated pseudo-masks guide the model to attend to diagnostically important anatomical regions, facilitating a systematic understanding of the overall scan context. Third, quantitative lesion information, including size, diameter, and spatial location, is encoded as structured textual prompts, enabling context-aware and clinically informed report generation. To enable rigorous evaluation, we validate our framework on multi-institutional structured report generation benchmarks. Experimental results demonstrate that MedRegion-CT achieves state-of-the-art performance, outperforming existing approaches in both linguistic quality and clinical accuracy. All code is publicly available at: https://github.com/babbu3682/MedRegion-CT.
\end{abstract}

\section{Introduction}
\label{section:introduction}

\begin{figure}[tb]
  \centering
  \vspace{-3mm}
  \includegraphics[width=0.45\textwidth]{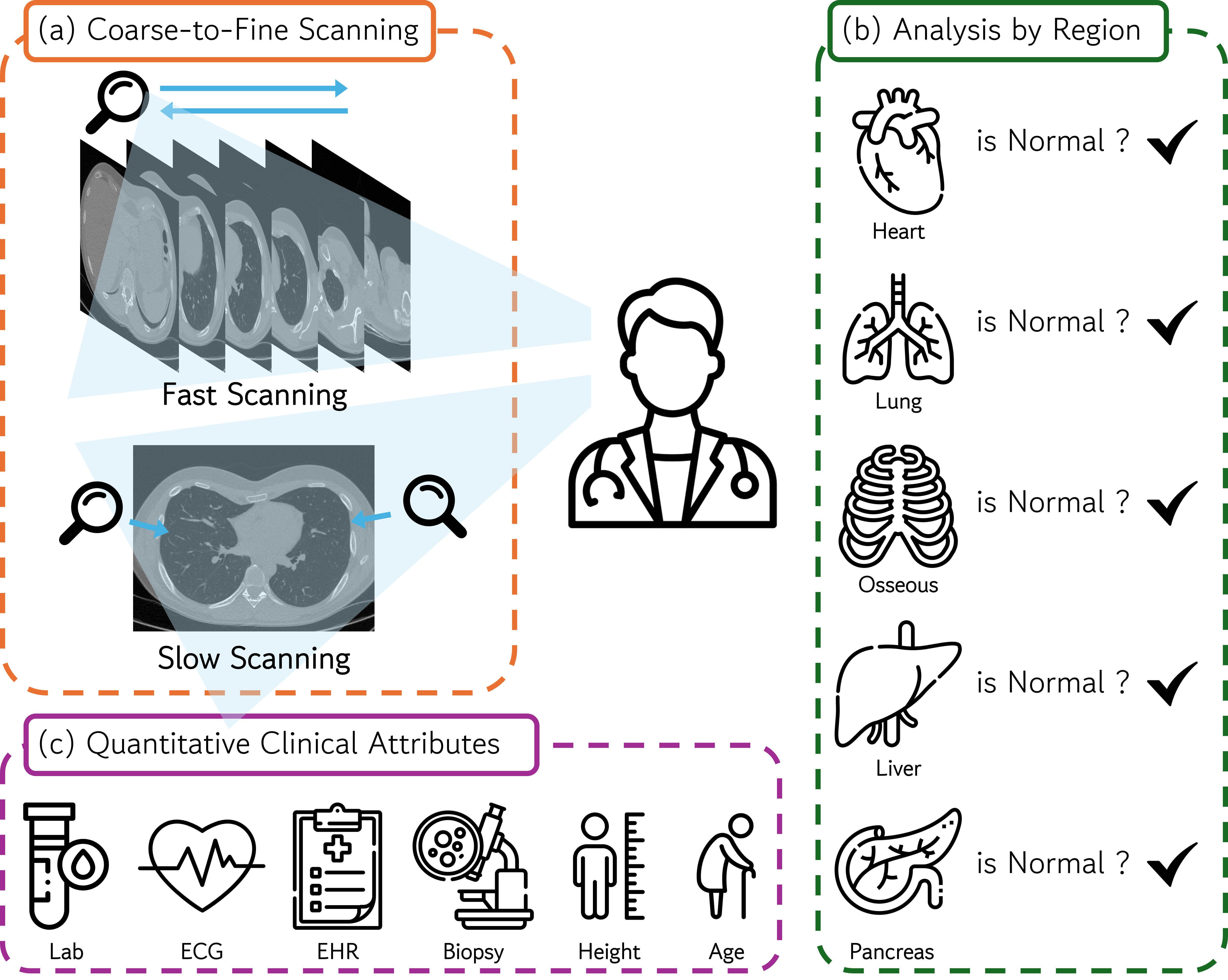}
  \caption{Comprehensive radiological workflow for clinical diagnosis. (a) Coarse-to-Fine Scanning: fast and slow scanning techniques for preliminary image review. (b) Analysis by Region: systematic assessment of major organ systems. (c) Quantitative Clinical Attributes: consideration of various patient-specific attributes for comprehensive clinical assessment.}
   \label{fig1}
   \vspace{-3mm}
\end{figure}

Computed tomography (CT) is a fundamental imaging modality widely used in clinical diagnosis, enabling detailed assessment of anatomical structures and pathological conditions \cite{hussain2022modern}. However, interpreting large volumes of CT slices and composing radiology reports remains a time-intensive process that imposes a substantial workload on radiologists \cite{goergen2013evidence, Claessens2015EarlyCC}. Consequently, automated systems that improve diagnostic efficiency while reducing clinical burden are highly desirable.

Recent advances in multimodal large language models (MLLMs) have achieved remarkable progress in complex multimodal reasoning tasks through strong language understanding capabilities and instruction tuning \cite{chen2024next}. In particular, recent open-source 3D medical MLLMs \cite{wu2023towards, bai2024m3d, xin2025med3dvlm, hamamci2024developing, shi2025medm} have demonstrated promising capabilities in jointly interpreting medical images and clinical information, supporting applications such as report generation and visual question answering.

Despite these advances, existing MLLMs primarily rely on global volume-level representations, which are insufficient for capturing fine-grained regional characteristics critical for accurate report generation. Advancing CT report generation therefore requires models that can effectively extract clinically meaningful regional information and integrate such localized evidence into the report generation process.

Radiologists initially review CT scans in a slice-wise manner to obtain a rapid holistic assessment, followed by focused examination of slices where lesions and organs are clearly visible for precise diagnosis (see Fig.~\ref{fig1}-(a)). Inspired by this workflow, we revisit the SlowFast strategy \cite{feichtenhofer2019slowfast} and introduce a Region-based SlowFast Tokenizer to steer the model's focus toward clinically meaningful regions. Guided by a holistic foreground mask of major organs, global fast tokens use coarse-grained aggregation to capture volumetric context with a low token count per slice, while regional slow tokens employ fine-grained aggregation to extract detailed pathological information with a higher token count. By targeting diagnostically salient regions, this dual-pathway approach yields clinically grounded representations optimized for 3D chest CT.

Subsequently, radiologists perform systematic organ-focused analysis to obtain detailed diagnostic evidence (see Fig.~\ref{fig1}-(b)). To emulate this process, we propose a Mask-Driven Visual Extractor. By leveraging pseudo-masks from a universal segmentation model, it extracts region-aware visual tokens of key organs. These tokens are subsequently integrated into the large language model (LLM) to foster a systematic understanding of anatomical structures.

In addition, diagnostic decisions are often supported by quantitative clinical attributes (see Fig.~\ref{fig1}-(c)). Accordingly, lesion pseudo-masks are processed using a deterministic algorithm to extract lesion attributes, including volume, diameter, and spatial location. These attributes are incorporated as structured textual prompts, enabling more precise and clinically informative report generation.

To facilitate rigorous evaluation, we construct multi-institutional structured report generation benchmarks, where each report is standardized into six clinically relevant, organ-specific sections.

In summary, we propose MedRegion-CT, an integrative framework for clinically precise CT report generation. Extensive experiments demonstrate that MedRegion-CT achieves state-of-the-art performance in structured report generation, producing reports with improved diagnostic accuracy and linguistic quality. Our contributions are as follows:

\begin{itemize}
   \item We introduce a \textbf{Region-based SlowFast Tokenizer} that constrains the model's focus to salient anatomical structures by utilizing a holistic foreground mask of major organs, enabling clinically-grounded learning of comprehensive and fine-grained 3D visual representations for CT report generation.
   \item We propose a \textbf{Mask-Driven Visual Extractor} that leverages pseudo masks of major organs to extract region-aware visual tokens, steering multimodal large language models to attend to major organs.
   \item We introduce a \textbf{Lesion Attribute Extractor} that derives quantitative lesion attributes, including volume, diameter, and spatial location, from pseudo-masks and encodes them as structured textual prompts to enhance clinically informative report generation.
   \item We construct a \textbf{Structured Chest CT Report Generation Benchmark} based on multi-institutional datasets and demonstrate state-of-the-art performance through extensive quantitative and qualitative evaluations across both linguistic and clinical assessment metrics.
\end{itemize}

\section{Related Work}
\label{section:related_work}

\subsubsection{3D CT Visual Encoding for Medical MLLMs.}
In the 3D medical MLLM domain, the extraction of representative features remains a significant challenge. In the visual encoding stage, compared to the 2D MLLM domain, the 3D MLLM domain faces constraints in transfer learning due to the absence of robust and well-optimized encoders pretrained on large-scale datasets. Moreover, 3D medical images inherently produce numerous vision tokens or features, resulting in substantial computational costs and contextual limitations for LLMs.

To address these challenges, existing studies generally adopt two primary strategies for 3D visual encoding. The first approach combines 3D vision transformer (ViT) encoders with token compression mechanisms. Recent works \cite{wu2023towards, bai2024m3d, hamamci2024developing} employ pretrained 3D ViT with input-level downsampling and compression modules such as attention pooling, which utilizes a small set of learnable latent queries, or spatial pooling in latent feature space to reduce token redundancy and computational cost. While this paradigm is intuitive and computationally efficient, aggressive token compression may inevitably discard fine-grained regional information that is critical for precise medical report generation.

The second approach adopts slice-wise 2D feature extraction followed by inter-slice aggregation. For instance, recent models aggregate slice-level representations using Z-formers \cite{lee2024read}, dual-encoding architectures \cite{shi2024med}, or attention mechanisms \cite{shi2025medm}. Benefiting from 2D vision encoders \cite{eslami2023pubmedclip, perez2025exploring} pretrained on large-scale datasets, such approaches effectively preserve slice-level details and inherit robust feature representations. However, such sequential 2D processing may overlook subtle spatial dependencies across the depth axis.

Recent advances in video recognition have demonstrated the effectiveness of the SlowFast dual-frame-rate strategy \cite{feichtenhofer2019slowfast}, which simultaneously models global temporal context and fine-grained motion information while maintaining computational efficiency \cite{Huang2024LITALI, Xu2024SlowFastLLaVAAS}. Inspired by this paradigm, we reinterpret the SlowFast strategy for 3D CT report generation and propose a Region-based SlowFast Tokenizer.

\subsubsection{Region-Focused Medical MLLMs.}
Several studies \cite{Zhang2023GPT4RoIIT, Ma2024GromaLV, Guo2024RegionGPTTR} in general domains have explored the incorporation of local features into LLMs, demonstrating the potential of region-level understanding. Even in the medical domain, radiology reports describe multiple anatomical regions simultaneously; therefore, recent studies have moved beyond conventional approaches relying on global-level representations, incorporating region-specific interpretations.

To identify clinically significant anatomical regions, several studies \cite{huang2024refer, zhou2025medversageneralistfoundationmodel, wang2025interpretablebilingualmultimodallarge} have constructed refer-and-ground instruction-following datasets that incorporate bounding box coordinates for model fine-tuning. Notably, MAIRA-2 \cite{Bannur2024MAIRA2GR} constructs a grounded radiology reporting dataset by meticulously annotating spatial locations corresponding to each CXR finding. This approach enables more precise and finding-specific interpretations. 

Several studies \cite{Tanida2023InteractiveAE, gao2024anatomy} have focused on leveraging region-of-interest (ROI) aligned features and integrating them into transformer-based language decoders to enhance radiology report generation. In particular, RGRG \cite{Tanida2023InteractiveAE} employs a Faster R-CNN detector to extract anatomical region-level features from chest X-rays, enabling more fine-grained and clinically consistent descriptions.


Recent approaches integrate pixel-level features into language decoders using pseudo-masks. Specifically, MAIRA-SEG \cite{sharma2024maira} extracts segmentation tokens from the pseudo-masks of major organs in chest X-ray images. Meanwhile, Reg2RG \cite{Chen2024LargeLM} utilizes a universal segmentation model to extract anatomical masks from chest CT images, integrating the resulting local features with global context for a cohesive contextual understanding.

However, comprehensive frameworks for 3D medical MLLMs that effectively leverage anatomical mask integration remain limited. To address this gap, we introduce a novel framework that leverages universal segmentation models to generate pseudo-masks of key anatomical structures, effectively integrating region-specific features via these pseudo-masks.

\subsubsection{Guided Textual Prompts for Reliable Medical MLLMs.}
Recent studies \cite{Li2023EvaluatingOH, Tong2024EyesWS} have raised concerns regarding the reliability of visual representations in MLLMs, as these models may inaccurately encode visual information, failing to answer even elementary visual questions. To address these limitations, several recent studies have developed text prompting approaches. Text prompting is a natural language processing (NLP) technique developed to transform inputs into textual templates, thereby improving language models' generalization capabilities and task adaptability, which has recently been systematized as guidelines for enhancing specific task performance \cite{liu2023pre}.


Recent studies have improved model reliability by transforming clinical and visual attributes into textual prompts. For instance, diagnostic results \cite{Jin2023PromptMRGDP}, disease classifications from CT images \cite{Chen2024DiaLLaMATL}, quantifiable morphological data \cite{Yeh2024InsightAM}, and voxel-level annotations \cite{Bassi2025RadGPTC3} have been successfully encoded as text prompts to guide LLMs. Inspired by these studies, we extract additional clinical information from pseudo-masks and integrate it as textual prompts to the LLM to generate more reliable medical reports.

\section{Method}
\label{section:method}

\begin{figure}[tb]
  \centering
  \vspace{-3mm}
  \includegraphics[width=\textwidth]{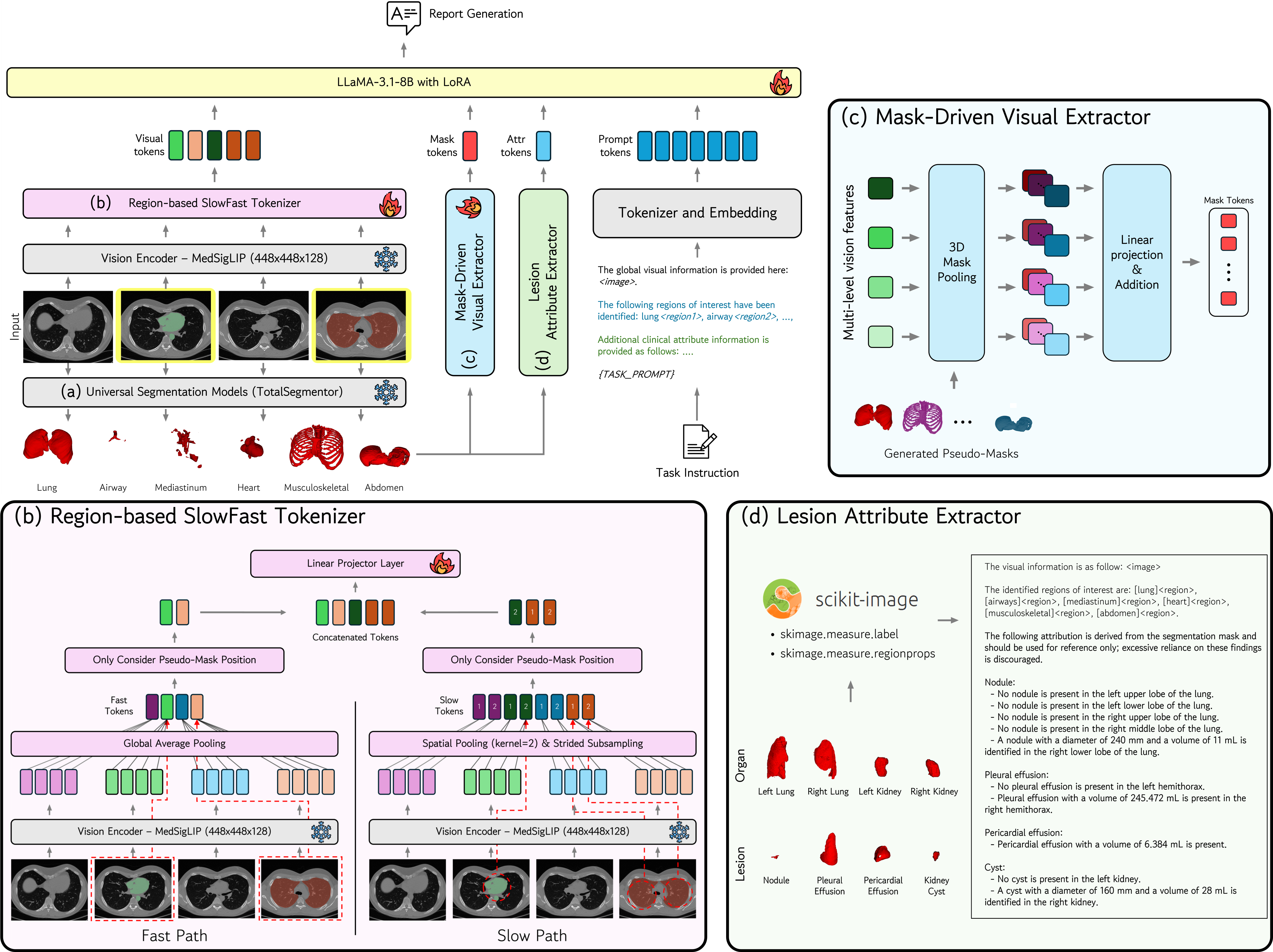}
   \caption{Overview of \textbf{MedRegion-CT}. (a) Pseudo-masks for six major organs and lesions are generated using a universal segmentation model. (b) Slice visual features are extracted slice-wise and aggregated into 3D visual tokens via a Region-based SlowFast Tokenizer. (c) A Mask-Driven Visual Extractor fuses multi-scale features and pseudo-masks into segmentation tokens. (d) A Lesion Attribute Extractor derives clinical attributes from pseudo-masks, producing attribute tokens for specific lesions such as lung nodules, pleural effusion, pericardial effusion, and kidney cyst. All tokens are input to an LLM, guided by instruction prompts, for report generation.}
   \label{fig2}
   \vspace{-3mm}
\end{figure}

The overall architecture of MedRegion-CT is illustrated in Fig.~\ref{fig2}. Initially, pseudo-masks for six predefined major clinical regions---including the lungs, airway, mediastinum, heart, musculoskeletal system, and abdomen---are generated from CT scans using a universal segmentation model (see Fig.~\ref{fig2}-(a)). Subsequently, slice-level visual features are extracted by processing 3D CT images slice-wise through a pre-trained 2D vision encoder (MedSigLIP \cite{sellergren2025medgemma}), which has been optimized through large-scale pre-training on millions of CT slices paired with their corresponding radiology reports. The extracted slice features are then transformed into 3D visual tokens, $T_{vision}$, via a Region-based SlowFast Tokenizer. This tokenizer integrates both global and regional tokens, each extracted based on the positions of an integrated multi-organ foreground mask (see Fig.~\ref{fig2}-(b)). This approach provides an efficient compressed representation that preserves both inter-slice and intra-slice information while emphasizing key anatomical regions.

Multi-scale visual features and pseudo-masks are then fed into a Mask-Driven Visual Extractor to generate mask tokens, $T_{mask}$ (see Fig.~\ref{fig2}-(c)). This mechanism enables focused analysis of anatomical structures. Additionally, a Lesion Attribute Extractor, incorporating deterministic algorithms, is applied to the pseudo-masks to extract quantitative medical information, thereby generating lesion-specific attribute tokens, $T_{attr}$ (see Fig.~\ref{fig2}-(d)). This ensures the generation of reliable reports regarding clinically relevant organs and lesions. 

To ensure standardized and clinically consistent outputs, we adopt a structured report format as the learning objective. This report is partitioned into six key anatomical sections, providing a comprehensive evaluation of the major clinical regions. By training the model on these structured labels, we explicitly guide the LLM to generate organized and domain-specific medical findings. Finally, $T_{vision}$, $T_{mask}$, and $T_{attr}$ are concatenated and fed into the LLM along with an instruction prompt $I$. The model then generates a structured medical report $R$, which can be formulated as follows:
\begin{equation}
  R = \text{LLM}(T_{vision}, T_{mask}, T_{attr}, I)
  \label{eq:report_gen}
\end{equation}
This strategy facilitates the generation of accurate, clinically grounded, and well-organized medical reports focused on the six predefined anatomical regions. The following subsections provide detailed explanations of each component.

\subsection{Region-based SlowFast Tokenizer}
\label{3_1}
While processing 3D CT images slice-wise through 2D encoders typically yields $D \times T$ tokens---where $D$ denotes the number of axial slices and $T$ represents the number of spatial tokens per slice---the direct input of such high-dimensional features into an LLM is computationally prohibitive. Existing SlowFast-based approaches \cite{Huang2024LITALI, Xu2024SlowFastLLaVAAS} attempt to mitigate this by combining spatially averaged tokens from all slices with tokens that are densely sampled from a subset of slices selected at uniform intervals. However, such uniform sampling strategies and the indiscriminate consideration of all slices are suboptimal for the 3D medical imaging domain.

To address this, we propose a Region-based SlowFast Tokenizer that leverages pseudo-masks to construct 3D visual tokens ($T_{vision}$) by isolating features that strictly correspond to predefined clinical regions. Let $f(i,j) \in \mathbb{R}^C$ denote the feature vector extracted from a 3D CT scan at the $i$-th depth slice ($i \in \{1, \dots, D\}$) and the $j$-th spatial token ($j \in \{1, \dots, T\}$), where $C$ is the channel dimension. We pair each feature with a binary indicator $m(i,j) \in \{0, 1\}$ derived from the foreground mask of major organs, where $m(i,j) = 1$ if $f(i,j)$ belongs to a predefined anatomical region, and $0$ otherwise. The tokenizer generates $T_{vision}$ through two specialized pathways: the Fast Pathway and the Slow Pathway.

\subsubsection{Fast Pathway (Global-Anatomical Context).} The Fast tokens, $T_{fast}$, capture the essential global context of the relevant anatomy. Instead of applying spatial average pooling to every depth slice indiscriminately, we first use the foreground mask to retain only organ-bearing slices, i.e., slices where the foreground mask of major organs is active. For each selected slice, we aggregate its slice-level representation by average pooling over the $T$ spatial tokens:
\begin{equation}
T_{fast} = \left\{ \text{AvgPool}_j(\{f(i,j)\}, kernel=T) \mid \textstyle\sum_{j} m(i, j) > 0 \right\}.
\end{equation}
This filters out non-informative background-only slices while preserving the global characteristics of each organ-bearing slice.

\subsubsection{Slow Pathway (Local-Detailed Context).} The Slow tokens, $T_{slow}$, preserve high-fidelity intra-slice details through region-guided spatial pooling. We harvest tokens strictly within the designated coordinates using masks with strided subsampling, and apply 2D spatial pooling with a kernel of 2:
\begin{equation}
T_{slow} = \left\{ \text{AvgPool}_j(\{f(i,j)\}, kernel=2) \mid m(i, 2j) = 1 \right\}.
\end{equation}
This focuses the model's capacity on granular pathological nuances while discarding the redundant overhead of non-clinical areas.

Finally, the tokens from both pathways are concatenated to form the final 3D visual representation $T_{vision}$. This dual-pathway strategy prioritizes clinically relevant volumetric features. By leveraging key organ segmentation masks, we extract dynamic, patient-specific tokens, significantly reducing the total sequence length to a manageable size while ensuring the LLM receives concentrated diagnostic information.

\subsection{Mask-Driven Visual Extractor}
\label{3_2}
To enable focused analysis of specific anatomical structures, we incorporate a Mask-Driven Visual Extractor based on the Osprey framework \cite{Yuan2023OspreyPU}. While the Osprey employs explicit location tokens, extending them to 3D CT volumes incurs prohibitive computational costs. Therefore, we omit these tokens to maintain efficiency and extend its mask-driven approach with a Volumetric Mask Pooling mechanism to effectively capture dense anatomical information across the entire volume.

Building upon the visual encoder, we extract multi-level slice visual features from the 6th, 12th, 18th, and 24th layers of MedSigLIP and apply our Mask Pooling module independently to each layer to distill these dense features into region-specific representations using key anatomical region masks. Specifically, for a given key anatomical region $o$, we adapt its corresponding binary pseudo-mask $m(i,j,o) \in \{0, 1\}$ to the feature resolution through trilinear interpolation. The layer-specific mask token, $k_{l}(o)$, is computed by pooling the features $f_{l}$ from layer $l$ using the provided mask:
\begin{equation}
k_{l}(o) = \sum_{(i,j) \in \Omega} f_l(i,j) \cdot \frac{m(i,j,o)}{\sum_{(i',j') \in \Omega} m(i',j',o) + \epsilon}
\end{equation}
where $\Omega$ defines the volumetric index set of size $D \times T$, $m(i,j,o)$ indicates the corresponding mask value for the target organ $o$, and $\epsilon$ is a small constant ($10^{-8}$).

Following the mask pooling for each layer, the resulting tokens $\{k_{l}(o)\}_{l=1}^4$ are passed through independent linear projectors, $\phi_l$. The projected tokens are then aggregated via element-wise summation. Finally, this aggregated representation is passed through a final linear projector, $\psi$, to map the fused features into the embedding space of the LLM:
\begin{equation}
T_{mask} = \psi \left( \sum_{l=1}^4 \phi_l(k_{l}(o)) \right)
\end{equation}

This hierarchical fusion strategy ensures that the final representation $T_{mask}$ incorporates both fine-grained anatomical details and high-level semantic context. By representing each region with a single condensed mask token, we minimize prompt length and ensure scalability for multi-organ analysis within the fixed context length of the LLM.

The extracted mask tokens are injected into the LLM at fixed prompt positions designated by special \texttt{<region>} tokens. This structured integration allows the LLM to focus specifically on designated anatomical structures, facilitating consistent and region-enhanced clinical report generation.

\subsection{Lesion Attribute Extractor}
\label{3_3}
Transmitting sufficient visual features to LLMs remains challenging, as 3D CT volumes are often heavily downsampled or cropped due to computational constraints, potentially leading to the loss of fine-grained morphological details. Relying solely on a vision encoder thus poses significant limitations for accurate diagnosis. To address this limitation, we integrate explicit descriptions of localized lesion characteristics into textual prompts, enabling the model to utilize these quantitative attributes that are crucial for determining disease states. Such characteristics—including volume, maximum diameter, and anatomical location—serve as vital diagnostic and prognostic biomarkers, particularly for assessing the malignancy risk of pulmonary nodules \cite{macmahon2017guidelines}.

Rather than relying on implicit features, we explicitly compute the geometric properties of lesions in the voxel space. Let $\mathcal{M}_{loc} \in \{0, 1\}^{D \times H \times W}$ and $\mathcal{M}_{les} \in \{0, 1\}^{D \times H \times W}$ denote the binary masks for an anatomical guidance region (e.g., right upper lobe) and a target lesion class (e.g., lung nodule), respectively. To isolate lesions strictly within a specific anatomical region, we compute their intersection and apply the largest connected component operator $\Phi$ to filter out segmentation noise:
\begin{equation}
    \mathcal{M}_{target} = \Phi(\mathcal{M}_{loc} \odot \mathcal{M}_{les})
\end{equation}
For the isolated instance $\mathcal{M}_{target}$, the absolute volume $V$ (in mL) and the maximum bounding-box diameter $d_{max}$ (in mm) are derived as:
\begin{equation}
    V = \left( \sum_{x,y,z} \mathcal{M}_{target}(x,y,z) \right) \times (s_x s_y s_z)/10^3, \quad d_{max} = \max_{i \in \{x,y,z\}} (L_i \times s_i)
\end{equation}
where $s_x, s_y, s_z$ represent the normalized voxel spacing along each axis, and $L_i$ denotes the voxel length of the bounding box of $\mathcal{M}_{target}$ along axis $i \in \{x,y,z\}$.

The specific types of detectable lesions are inherently determined by the universal segmentation model employed. Utilizing TotalSegmentator \cite{wasserthal2023totalsegmentator} for initial segmentation, alongside SimpleITK and scikit-image for morphological post-processing, we extract quantitative profiles for critical lesions including pulmonary nodules, pleural/pericardial effusions, and renal cysts. This metadata is restructured into a textual format (see Supplementary Algorithm \ref{supple_alg1}) and encoded as attribute tokens $T_{attr}$ for the LLM, as illustrated in Fig.~\ref{fig2}-(d). This approach contributes to generating consistent and reliable region-focused reports while mitigating the variability inherent in human observations.

\subsection{Structured Report Generation Benchmark}
\label{3_4}
Publicly available radiology reports often lack systematic organization, hindering both supervised learning and consistent clinical evaluation. To address this, we establish a structured chest CT report generation benchmark by reorganizing unstructured data into a standardized format. We divide the anatomical field of view into six primary clinical regions (e.g., lung parenchyma, large airway, and mediastinum, among others). This categorization reflects the systematic approach employed by radiologists, ensuring thorough evaluation of relevant pathology within each region and enhancing overall report quality and consistency \cite{Marcovici2014JournalCS, Cereser2024AssessingTI}. Detailed specifications for these categories and their constituent sub-regions are provided in Supplementary Tab.~\ref{supplementary_tab2}.

To construct high-quality training labels, we utilized the RadGenome-Chest CT dataset \cite{zhang2024radgenome}, which provides multi-granularity grounded reports. While this dataset decomposes findings into numerous fine-grained structures, it lacks a unified representation for the six major clinical regions essential for comprehensive chest evaluation. Therefore, we developed a multi-stage NLP pipeline using the DSPy \cite{khattab2023dspy} framework and the Llama-3.3-70B-Instruct \cite{grattafiori2024llama} model to selectively aggregate and refine these fragmented sentences. This pipeline systematically performs: (i) clinical cleansing to expand medical abbreviations; (ii) temporal refinement to focus on current findings by filtering out longitudinal comparison noise; and (iii) anatomical merging to integrate sub-region findings into the predefined clinical hierarchy.

To enable consistent evaluation, we applied this NLP pipeline to both the internal RadGenome-Chest CT dataset and the external Asan Medical Center (AMC) dataset. Reorganizing each ground-truth report into this unified format effectively constructs a standardized report generation benchmark for both in-domain and cross-institutional settings. These benchmarks facilitate systematic assessments of model performance across linguistic and clinical metrics, establishing a reliable foundation for automated structured reporting in 3D medical imaging.


\section{Experiments}
\label{section:experiments}


\begin{table}[tb] 
\caption{Comparisons with state-of-the-art methods on the RadGenome-Chest CT and AMC structured report generation benchmarks. Brackets denote the 95\% confidence interval (CI). \textbf{Bold} and \underline{underlined} values indicate the best and second-best results, respectively.}
\label{tab1}
\vspace{-3mm}
\centering 
\resizebox{\textwidth}{!}{
\begin{tabular}{l ccc ccc} 
\toprule
\multirow{2}{*}{\textbf{Method}} & \multicolumn{3}{c}{\textbf{NLG Metrics}} & \multicolumn{3}{c}{\textbf{LM-based Metrics}} \\
\cmidrule(lr){2-4} \cmidrule(lr){5-7}
& \textbf{BLEU} & \textbf{ROUGE} & \textbf{METEOR} & \textbf{GREEN} & \textbf{CRG} & \textbf{GPT-CA} \\
\midrule
\multicolumn{7}{c}{\textit{Structured RadGenome-Chest CT (Internal Dataset, n=1551)}} \\
\midrule
RadFM \cite{wu2023towards} & 0.3153 & 0.3732 & 0.4894 & 0.3030 & 0.3477 & 0.2852 \\
& {\scriptsize [0.3072, 0.3237]} & {\scriptsize [0.3667, 0.3798]} & {\scriptsize [0.4825, 0.4966]} & {\scriptsize [0.2914, 0.3142]} & {\scriptsize [0.3447, 0.3508]} & {\scriptsize [0.2771, 0.2936]} \\
\addlinespace[0.5ex]
M3D \cite{bai2024m3d} & 0.3177 & 0.3973 & 0.4938 & \underline{0.4013} & 0.3717 & 0.3586 \\
& {\scriptsize [0.3083, 0.3277]} & {\scriptsize [0.3889, 0.4058]} & {\scriptsize [0.4860, 0.5018]} & {\scriptsize [0.3883, 0.4146]} & {\scriptsize [0.3683, 0.3751]} & {\scriptsize [0.3479, 0.3694]} \\
\addlinespace[0.5ex]
MedM-VL \cite{shi2025medm} & 0.3026 & 0.3854 & 0.4811 & 0.3970 & 0.3426 & 0.3168 \\
& {\scriptsize [0.2920, 0.3116]} & {\scriptsize [0.3775, 0.3934]} & {\scriptsize [0.4734, 0.4892]} & {\scriptsize [0.3856, 0.4084]} & {\scriptsize [0.3408, 0.3445]} & {\scriptsize [0.3066, 0.3273]} \\
\addlinespace[0.5ex]
Med3DVLM \cite{xin2025med3dvlm} & \underline{0.3269} & \underline{0.3991} & \underline{0.5029} & 0.3881 & \underline{0.3866} & \underline{0.3733} \\
& {\scriptsize [0.3174, 0.3368]} & {\scriptsize [0.3905, 0.4078]} & {\scriptsize [0.4946, 0.5111]} & {\scriptsize [0.3738, 0.4024]} & {\scriptsize [0.3828, 0.3907]} & {\scriptsize [0.3622, 0.3846]} \\
\addlinespace[0.5ex]
CT-CHAT \cite{hamamci2024developing} & 0.3081 & 0.3900 & 0.4831 & 0.3966 & 0.3425 & 0.3265 \\
& {\scriptsize [0.2980, 0.3175]} & {\scriptsize [0.3815, 0.3987]} & {\scriptsize [0.4749, 0.4917]} & {\scriptsize [0.3843, 0.4092]} & {\scriptsize [0.3408, 0.3443]} & {\scriptsize [0.3155, 0.3374]} \\
\addlinespace[0.5ex]
\textbf{Ours} & \textbf{0.3435} & \textbf{0.4225} & \textbf{0.5205} & \textbf{0.4555} & \textbf{0.3959} & \textbf{0.4003} \\
& {\scriptsize [0.3325, 0.3534]} & {\scriptsize [0.4132, 0.4318]} & {\scriptsize [0.5118, 0.5291]} & {\scriptsize [0.4411, 0.4697]} & {\scriptsize [0.3919, 0.4003]} & {\scriptsize [0.3884, 0.4124]} \\
\midrule
\multicolumn{7}{c}{\textit{Structured AMC (External Dataset, n=500, Zero-shot Inference)}} \\
\midrule
RadFM \cite{wu2023towards} & \textbf{0.1464} & \textbf{0.1928} & \textbf{0.4790} & 0.0977 & 0.3476 & 0.2018 \\
& {\scriptsize [0.1424, 0.1502]} & {\scriptsize [0.1889, 0.1969]} & {\scriptsize [0.4730, 0.4851]} & {\scriptsize [0.0802, 0.1164]} & {\scriptsize [0.3427, 0.3527]} & {\scriptsize [0.1880, 0.2161]} \\
\addlinespace[0.5ex]
M3D \cite{bai2024m3d} & 0.1185 & 0.1565 & 0.4457 & 0.1102 & 0.3725 & 0.2243 \\
& {\scriptsize [0.1137, 0.1218]} & {\scriptsize [0.1537, 0.1593]} & {\scriptsize [0.4399, 0.4515]} & {\scriptsize [0.0901, 0.1317]} & {\scriptsize [0.3666, 0.3787]} & {\scriptsize [0.2104, 0.2389]} \\
\addlinespace[0.5ex]
MedM-VL \cite{shi2025medm} & 0.1235 & 0.1543 & 0.4430 & 0.1115 & 0.3472 & 0.1963 \\
& {\scriptsize [0.1216, 0.1254]} & {\scriptsize [0.1517, 0.1569]} & {\scriptsize [0.4371, 0.4489]} & {\scriptsize [0.0909, 0.1336]} & {\scriptsize [0.3436, 0.3508]} & {\scriptsize [0.1792, 0.2143]} \\
\addlinespace[0.5ex]
Med3DVLM \cite{xin2025med3dvlm} & 0.1136 & 0.1557 & 0.4400 & \underline{0.1138} & \underline{0.3791} & \underline{0.2366} \\
& {\scriptsize [0.1106, 0.1166]} & {\scriptsize [0.1525, 0.1590]} & {\scriptsize [0.4338, 0.4462]} & {\scriptsize [0.0933, 0.1361]} & {\scriptsize [0.3725, 0.3863]} & {\scriptsize [0.2215, 0.2527]} \\
\addlinespace[0.5ex]
CT-CHAT \cite{hamamci2024developing} & 0.1173 & 0.1559 & 0.4410 & \underline{0.1138} & 0.3487 & 0.2053 \\
& {\scriptsize [0.1064, 0.1261]} & {\scriptsize [0.1529, 0.1589]} & {\scriptsize [0.4345, 0.4474]} & {\scriptsize [0.0926, 0.1363]} & {\scriptsize [0.3449, 0.3525]} & {\scriptsize [0.1870, 0.2247]} \\
\addlinespace[0.5ex]
\textbf{Ours} & \underline{0.1275} & \underline{0.1572} & \underline{0.4467} & \textbf{0.1572} & \textbf{0.3822} & \textbf{0.2642} \\
& {\scriptsize [0.1150, 0.1354]} & {\scriptsize [0.1544, 0.1601]} & {\scriptsize [0.4412, 0.4521]} & {\scriptsize [0.1305, 0.1804]} & {\scriptsize [0.3749, 0.3903]} & {\scriptsize [0.2470, 0.2825]} \\
\bottomrule
\end{tabular}%
} 
\vspace{-3mm}
\end{table}

\subsection{Datasets and Metrics}
\subsubsection{Datasets.} We utilize two structured report generation benchmarks to comprehensively evaluate our model: the internal RadGenome-Chest CT benchmark for assessing in-domain performance, and the external AMC benchmark for evaluating zero-shot generalization under cross-institutional domain shift. Specifically, the RadGenome-Chest CT follows the official split, comprising 23,865 training/validation and 1,551 testing scans with corresponding multi-granularity reports. For the external AMC dataset, we incorporate non-contrast chest CT scans and paired reports collected from 4,014 patients between 2010 and 2021, from which a subset of 500 cases was randomly sampled. The retrospective use of the AMC dataset was approved by the institutional review board, and the requirement for informed consent was waived. All reports from both datasets were further processed into the structured format described in Sec.~\ref{3_4}.

\subsubsection{Metrics.} Evaluating generated radiology reports requires assessing both linguistic similarity and clinical accuracy. Our methodology incorporates two metric categories: 1) Natural Language Generation (NLG) metrics include BLEU \cite{papineni-etal-2002-bleu}, ROUGE \cite{lin-2004-rouge}, and METEOR \cite{Banerjee2005METEORAA}. 2) Language Model (LM) based metrics consist of CRG \cite{hamamci2025crg}, GREEN \cite{Ostmeier2024GREENGR}, and Clinical Accuracy (CA) \cite{hamamci2024developing} using the gpt-5-nano API. These metrics ensure comprehensive assessment across structural, linguistic, and clinical aspects essential for radiology reporting.

\subsubsection{Implementations.}
We implemented our framework using PyTorch and trained it on NVIDIA A100 GPUs. The input 3D CT volumes were standardized to a fixed resolution of $448 \times 448 \times 128$. We utilized the pre-trained MedSigLIP as the vision encoder and LLaMA-3.1-8B \cite{grattafiori2024llama} as the language model, applying LoRA \cite{Hu2021LoRALA} for parameter-efficient fine-tuning. The model was trained in two stages: initially aligning the multimodal connector and mask projector, followed by fine-tuning on the structured report dataset. Detailed hyperparameters and preprocessing configurations are provided in Supplementary Sec.~\ref{supplementary_sec1}.

\begin{figure}[tb]
\centering
\vspace{-3mm}
\includegraphics[width=\textwidth, trim=0 16 0 0, clip]{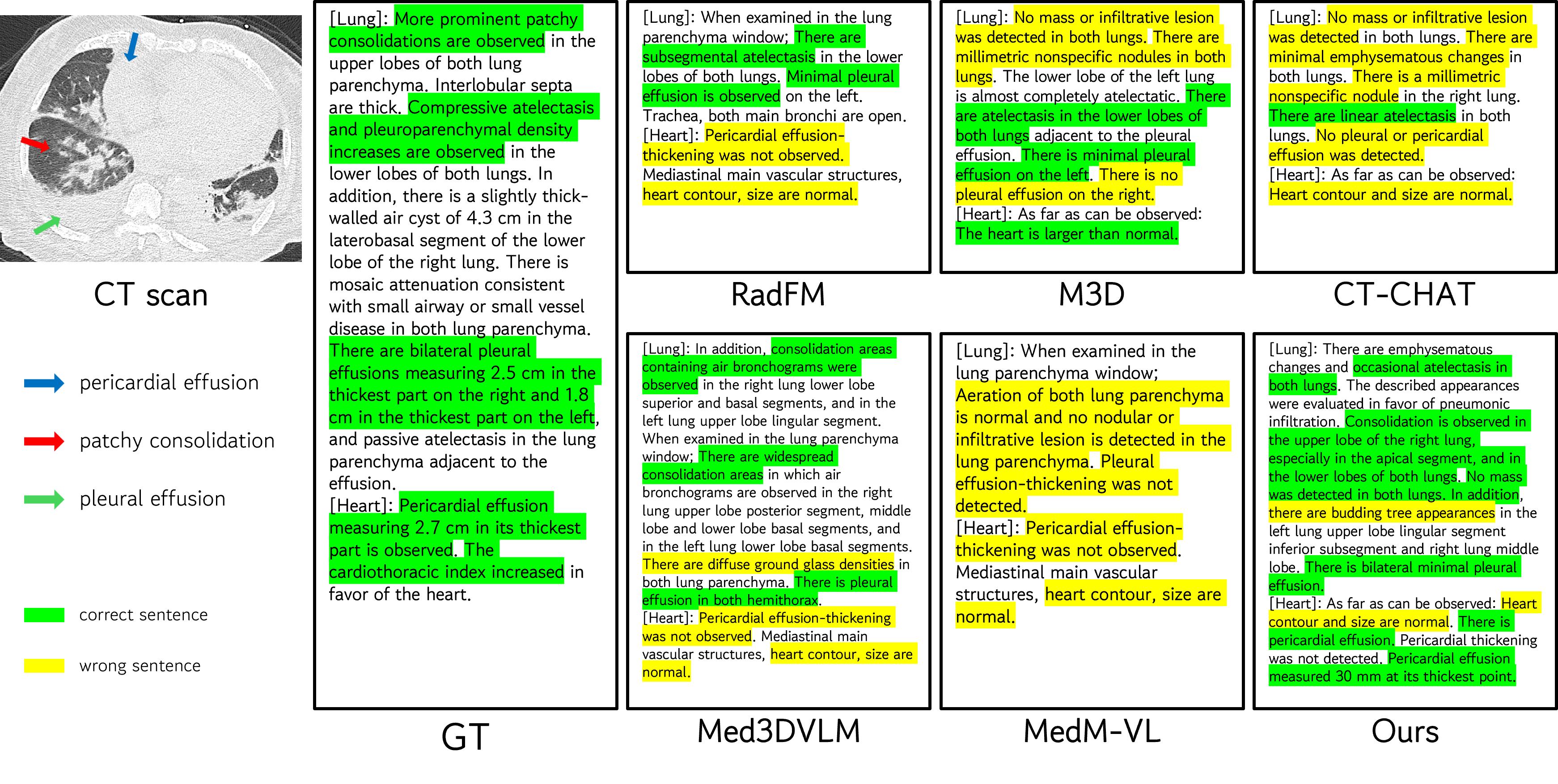}
\caption{Qualitative comparison of generated radiology reports for a patient exhibiting significant pericardial and pleural effusions with associated pulmonary consolidations. We compare our proposed MedRegion-CT against state-of-the-art methods. Text is color-coded to indicate findings consistent with the ground truth (\textcolor{green}{green}) 
and findings that are inconsistent with the ground truth or represent hallucinations (\textcolor{yellow}{yellow}). For clarity, only the lung and heart sections of the report are visualized. MedRegion-CT demonstrates superior performance in both diagnostic accuracy and localization performance.}
\label{fig3}
\vspace{-3mm}
\end{figure}

\subsection{Comparison with State-of-the-Art 3D MLLMs}
To evaluate our proposed framework, we compared it against five state-of-the-art 3D Medical MLLMs (RadFM \cite{wu2023towards}, M3D \cite{bai2024m3d}, CT-CHAT \cite{hamamci2024developing}, Med3DVLM \cite{xin2025med3dvlm}, and MedM-VL \cite{shi2025medm}) on the internal RadGenome-Chest CT and external AMC benchmarks. All baselines were initialized with their official weights and fine-tuned on our structured internal dataset to ensure optimal localized report generation. These models were then evaluated on the external AMC dataset without additional training to assess zero-shot generalization.

\subsubsection{Quantitative Results.} As summarized in Tab.~\ref{tab1}, MedRegion-CT consistently outperforms the baselines across all evaluation categories on the internal benchmark. Our model achieves the highest NLG metrics (BLEU: 0.3435, ROUGE: 0.4225, METEOR: 0.5205), indicating strong lexical and semantic alignment with ground-truth reports. Furthermore, it achieves the best LM-based clinical scores (GREEN: 0.4555, CRG: 0.3959, GPT-CA: 0.4003). While M3D and Med3DVLM show competitive results in specific metrics, they fall short of our model's overall performance. The substantial margin in clinical accuracy, particularly in GREEN, demonstrates that our approach more effectively captures essential diagnostic findings within the six predefined clinical regions (see Supplementary Tab.~\ref{supplementary_tab1} for detailed region-wise results). A further size-stratified nodule analysis in Supplementary Tab.~\ref{supplementary_tab3} confirms superior detection rates for mid-to-large nodules compared to existing baselines. Although performance decreases for small abnormalities ($\le 5\text{mm}^3$), this is primarily bounded by the false negatives of the upstream off-the-shelf segmentation model. 

Under zero-shot inference on the external AMC benchmark, MedRegion-CT demonstrates robust generalization against cross-institutional domain shifts. Although RadFM achieves the highest NLG scores, its lower LM-based clinical scores suggest that lexical-overlap metrics may favor institution-specific reporting styles over clinical faithfulness. In contrast, MedRegion-CT achieves the best LM-based performance (GREEN: 0.1572, CRG: 0.3822, GPT-CA: 0.2642) while maintaining the second-best NLG performance. This divergence highlights the limitation of surface-level lexical metrics in cross-institutional evaluations, where report phrasing and formatting vary substantially across hospitals.

Consistent with the results of the radiologist reader study (Supplementary Tab.~\ref{supplementary_tab4}), which highlight our model's strong ability to capture key findings and minimize hallucinations, MedRegion-CT preserves clinically relevant information even when diverging from the exact wording of reference reports. Conversely, models relying primarily on global features (e.g., RadFM, CT-CHAT) exhibit lower clinical reliability in structured reporting, likely due to the difficulty of localized feature extraction in complex 3D volumes. Overall, our framework generates clinically grounded reports with strong zero-shot generalization.

\subsubsection{Qualitative Results.} Fig.~\ref{fig3} illustrates a qualitative comparison of generated reports on the internal structured benchmark. The ground-truth report for this case describes bilateral consolidations accompanied by significant pericardial and pleural effusions. MedRegion-CT demonstrates superior diagnostic fidelity by accurately identifying these multi-regional pathologies. Notably, it correctly captures the upper-lobe consolidations and reflects the presence of effusions with high precision, closely aligning with the ground-truth (GT). 

In contrast, baseline models exhibit several clinical inaccuracies. RadFM and Med3DVLM miss the critical pericardial effusion, incorrectly reporting normal heart contours. M3D overlooks the primary parenchymal consolidations in favor of non-specific nodules, while MedM-VL provides an essentially normal assessment. These qualitative results further support the quantitative findings, confirming that MedRegion-CT effectively minimizes hallucinations and provides reliable clinical utility through precise regional localization.

\begin{table}[tb]
\caption{Ablation study of MedRegion-CT components on the RadGenome-Chest CT and AMC structured report generation benchmarks. Here, $R$ denotes our Region-based SlowFast Tokenizer, $Attr$ denotes our Lesion Attribute Extractor, and $Mask$ denotes our Mask-Driven Visual Extractor. LITA$_{enc.}$ and M3D$_{enc.}$ indicate the conventional SlowFast tokenization strategy \cite{Huang2024LITALI} and the pre-trained 3D ViT encoder from M3D \cite{bai2024m3d}, respectively.}
\label{tab2}
\vspace{-3mm}
\centering 
\resizebox{\textwidth}{!}{
    \begin{tabular}{l ccc ccc}
        \toprule
        \multirow{2}{*}{\textbf{Method}} & \multicolumn{3}{c}{\textbf{NLG Metrics}} & \multicolumn{3}{c}{\textbf{LM-based Metrics}} \\
        \cmidrule(lr){2-4} \cmidrule(lr){5-7}
        & \textbf{BLEU} & \textbf{ROUGE} & \textbf{METEOR} & \textbf{GREEN} & \textbf{CRG} & \textbf{GPT-CA} \\
        \midrule
        \multicolumn{7}{c}{\textit{Structured RadGenome-Chest CT (Internal Dataset, n=1551)}} \\
        \midrule
        $R$+$Attr$+$Mask$ & \textbf{0.3273} & \textbf{0.3988} & \textbf{0.5039} & \textbf{0.4025} & \textbf{0.3861} & \textbf{0.3839} \\
        & {\scriptsize [0.3161, 0.3364]} & {\scriptsize [0.3904, 0.4075]} & {\scriptsize [0.4960, 0.5121]} & {\scriptsize [0.3889, 0.4165]} & {\scriptsize [0.3821, 0.3901]} & {\scriptsize [0.3728, 0.3952]} \\
        \addlinespace[0.5ex]
        $R$+$Attr$ & 0.3271 & 0.3984 & 0.5034 & 0.4010 & 0.3846 & 0.3807 \\
        & {\scriptsize [0.3131, 0.3361]} & {\scriptsize [0.3898, 0.4070]} & {\scriptsize [0.4955, 0.5115]} & {\scriptsize [0.3876, 0.4148]} & {\scriptsize [0.3804, 0.3887]} & {\scriptsize [0.3728, 0.3952]} \\
        \addlinespace[0.5ex]
        $R$+$Mask$ & 0.3248 & 0.3973 & 0.4996 & 0.3937 & 0.3699 & 0.3587 \\
        & {\scriptsize [0.3148, 0.3343]} & {\scriptsize [0.3887, 0.4060]} & {\scriptsize [0.4916, 0.5076]} & {\scriptsize [0.3797, 0.4072]} & {\scriptsize [0.3664, 0.3736]} & {\scriptsize [0.3473, 0.3701]} \\
        \addlinespace[0.5ex]
        $R$      & 0.3152 & 0.3909 & 0.4912 & 0.3930 & 0.3613 & 0.3513 \\
        & {\scriptsize [0.3059, 0.3240]} & {\scriptsize [0.3827, 0.3992]} & {\scriptsize [0.4836, 0.4993]} & {\scriptsize [0.3794, 0.4068]} & {\scriptsize [0.3583, 0.3645]} & {\scriptsize [0.3400, 0.3629]} \\
        \addlinespace[0.5ex]
        M3D$_{enc.}$ & 0.3148 & 0.3917 & 0.4885 & 0.4009 & 0.3487 & 0.3312 \\
        & {\scriptsize [0.3048, 0.3238]} & {\scriptsize [0.3834, 0.4002]} & {\scriptsize [0.4804, 0.4969]} & {\scriptsize [0.3878, 0.4143]} & {\scriptsize [0.3463, 0.3513]} & {\scriptsize [0.3203, 0.3422]} \\
        \addlinespace[0.5ex]
        LITA$_{enc.}$ & 0.3106 & 0.3902 & 0.4904 & 0.3880 & 0.3562 & 0.3587 \\
        & {\scriptsize [0.3009, 0.3199]} & {\scriptsize [0.3822, 0.3983]} & {\scriptsize [0.4822, 0.4984]} & {\scriptsize [0.3801, 0.4078]} & {\scriptsize [0.3533, 0.3591]} & {\scriptsize [0.3477, 0.3700]} \\
        \midrule
        \multicolumn{7}{c}{\textit{Structured AMC (External Dataset, n=500, Zero-shot Inference)}} \\
        \midrule
        $R$+$Attr$+$Mask$ & \textbf{0.1137} & 0.1577 & \textbf{0.4514} & \textbf{0.1425} & \textbf{0.3772} & \textbf{0.2564} \\
        & {\scriptsize [0.1112, 0.1160]} & {\scriptsize [0.1548, 0.1607]} & {\scriptsize [0.4461, 0.4568]} & {\scriptsize [0.1188, 0.1678]} & {\scriptsize [0.3709, 0.3843]} & {\scriptsize [0.2404, 0.2731]} \\
        \addlinespace[0.5ex]
        $R$+$Attr$      & 0.1076 & 0.1570 & 0.4493 & 0.1365 & 0.3762 & 0.2530 \\
        & {\scriptsize [0.0963, 0.1151]} & {\scriptsize [0.1540, 0.1601]} & {\scriptsize [0.4438, 0.4549]} & {\scriptsize [0.1130, 0.1611]} & {\scriptsize [0.3698, 0.3838]} & {\scriptsize [0.2368, 0.2703]} \\
        \addlinespace[0.5ex]
        $R$+$Mask$      & 0.1128 & \textbf{0.1628} & 0.4207 & 0.1313 & 0.3616 & 0.2323 \\
        & {\scriptsize [0.1098, 0.1160]} & {\scriptsize [0.1595, 0.1662]} & {\scriptsize [0.4149, 0.4267]} & {\scriptsize [0.1084, 0.1552]} & {\scriptsize [0.3569, 0.3665]} & {\scriptsize [0.2160, 0.2494]} \\
        \addlinespace[0.5ex]
        $R$           & 0.1081 & 0.1588 & 0.4465 & 0.1280 & 0.3645 & 0.2244 \\
        & {\scriptsize [0.0913, 0.1229]} & {\scriptsize [0.1557, 0.1620]} & {\scriptsize [0.4406, 0.4522]} & {\scriptsize [0.1057, 0.1514]} & {\scriptsize [0.3584, 0.3707]} & {\scriptsize [0.2063, 0.2390]} \\
        \addlinespace[0.5ex]
        M3D$_{enc.}$     & 0.1126 & 0.1566 & 0.4450 & 0.1221 & 0.3576 & 0.2126 \\
        & {\scriptsize [0.1000, 0.1216]} & {\scriptsize [0.1537, 0.1595]} & {\scriptsize [0.4393, 0.4509]} & {\scriptsize [0.1010, 0.1443]} & {\scriptsize [0.3522, 0.3638]} & {\scriptsize [0.1966, 0.2296]} \\
        \addlinespace[0.5ex]
        LITA$_{enc.}$    & 0.1079 & 0.1578 & 0.4473 & 0.1264 & 0.3597 & 0.2241 \\
        & {\scriptsize [0.1053, 0.1105]} & {\scriptsize [0.1547, 0.1607]} & {\scriptsize [0.4420, 0.4528]} & {\scriptsize [0.1042, 0.1498]} & {\scriptsize [0.3547, 0.3657]} & {\scriptsize [0.2078, 0.2414]} \\
        \bottomrule
    \end{tabular}%
} 
\vspace{-3mm}
\end{table}

\subsection{Ablation Study}
To evaluate the contribution of each module, we conducted an ablation study using a 30\% training subset of the RadGenome-Chest CT dataset and validated zero-shot generalization on the AMC dataset. We compared our full configuration ($R$+$Attr$+$Mask$) against internal variants and external encoder baselines. Here, $R$, $Attr$, and $Mask$ denote the Region-based SlowFast Tokenizer, Lesion Attribute Extractor, and Mask-Driven Visual Extractor, respectively. LITA${enc.}$ \cite{Huang2024LITALI} and M3D${enc.}$ \cite{bai2024m3d} represent the conventional SlowFast strategy and the pre-trained 3D ViT encoder, respectively.

On the internal benchmark (Tab.~\ref{tab2}, top), our full model outperforms all variants across both NLG (BLEU: 0.3273, ROUGE: 0.3988, METEOR: 0.5039) and clinical metrics (GREEN: 0.4025, CRG: 0.3861, GPT-CA: 0.3839), demonstrating its ability to generate fluent and clinically grounded reports. The incremental addition of $Attr$ and $Mask$ to the $R$ baseline steadily improves performance. Notably, $Attr$ substantially enhances clinical consistency by integrating deterministic quantitative biomarkers. To further analyze our model's dependence on segmentation quality, Supplementary Tab.~\ref{supplementary_tab5} presents a mask perturbation stress test, showing that organ-mask distortions affect performance more strongly than lesion-mask distortions by weakening ROI guidance.

Zero-shot inference on the external AMC benchmark (Tab.~\ref{tab2}, bottom) validates the architecture's generalizability. The full model achieves the highest scores across most metrics, particularly in clinical evaluations (GREEN: 0.1425, CRG: 0.3772, GPT-CA: 0.2564). While $R$+$Mask$ marginally outperforms the full model in ROUGE (0.1628 vs. 0.1577), integrating $Attr$ is crucial for maintaining clinical accuracy on unseen distributions, evidenced by the distinct improvements in GREEN and GPT-CA. Finally, comparing our standalone regional tokenizer ($R$) with external encoders (M3D${enc.}$, LITA${enc.}$) demonstrates $R$'s superior transferability. This indicates that our region-prioritized architecture provides a more effective foundation for 3D feature extraction by concentrating the model's capacity on clinically salient anatomical areas, which directly translates to robust performance and generalization on external datasets.

Supplementary Fig.~\ref{supple_fig1} visualizes the diagnostic influence of each component. The full model ($R+Attr+Mask$) correctly identifies the hallmark ground-glass opacities (GGOs) in the peripheral subpleural area and establishes an accurate clinical diagnosis of viral pneumonia, closely matching the GT. When $Mask$ is removed ($R+Attr$), the model still captures the viral pneumonia diagnosis but shows less precision in describing the subpleural distribution. Conversely, removing $Attr$ ($R+Mask$) results in the identification of nodules and GGOs but introduces erroneous emphysematous changes not present in the GT. The baseline variants, including $R$, M3D$_{enc.}$, and LITA$_{enc.}$, fail significantly by either reporting normal aeration or identifying non-specific nodules while missing the primary GGO findings. This qualitative evidence confirms that our integrated approach is necessary to maintain both localization precision and diagnostic coherence across varying clinical scenarios.

\section{Conclusion}
\label{section:conclusion}
In this work, we presented \textbf{MedRegion-CT}, a novel MLLM framework that shifts the paradigm from global volumetric encoding to region-aware 3D representation for CT report generation. By integrating a Region-based SlowFast Tokenizer, a Mask-Driven Visual Extractor, and a Lesion Attribute Extractor, our model effectively emulates the hierarchical diagnostic workflow of radiologists. These innovations enable precise capture of fine-grained pathological details and quantitative clinical metadata that are often overlooked by conventional global-feature approaches. Extensive evaluation on our Structured Chest CT Report Generation Benchmarks demonstrates that MedRegion-CT establishes a new state-of-the-art, significantly outperforming existing 3D MLLMs in both linguistic fluency and clinical accuracy. Our results confirm that regional guidance is essential for minimizing hallucinations and ensuring reliable diagnostic reporting. We believe this region-centric approach provides a scalable and robust foundation for future 3D medical image interpretation.



\section*{Acknowledgements}
The authors would like to express their sincere gratitude to Jihyun Kim, Junseong Lee, and Jeong Min Song for their invaluable assistance with external data curation, as well as their comprehensive technical and experimental support throughout this work.

\clearpage
%
%
\bibliographystyle{splncs04}
\bibliography{main}

@String(CVPR  = {IEEE Conf. Comput. Vis. Pattern Recog.})

@String(CVPR  = {CVPR})

@article{bai2024m3d,
  title={M3d: Advancing 3d medical image analysis with multi-modal large language models},
  author={Bai, Fan and Du, Yuxin and Huang, Tiejun and Meng, Max Q-H and Zhao, Bo},
  journal={arXiv preprint arXiv:2404.00578},
  year={2024}
}

@article{hussain2022modern,
  title={Modern diagnostic imaging technique applications and risk factors in the medical field: a review},
  author={Hussain, Shah and Mubeen, Iqra and Ullah, Niamat and Shah, Syed Shahab Ud Din and Khan, Bakhtawar Abduljalil and Zahoor, Muhammad and Ullah, Riaz and Khan, Farhat Ali and Sultan, Mujeeb A},
  journal={BioMed research international},
  volume={2022},
  number={1},
  pages={5164970},
  year={2022},
  publisher={Wiley Online Library}
}

@article{goergen2013evidence,
  title={Evidence-based guideline for the written radiology report: Methods, recommendations and implementation challenges},
  author={Goergen, Stacy K and Pool, Felicity J and Turner, Tari J and Grimm, Jane E and Appleyard, Mark N and Crock, Carmel and Fahey, Michael C and Fay, Michael F and Ferris, Nicholas J and Liew, Susan M and others},
  journal={Journal of medical imaging and radiation oncology},
  volume={57},
  number={1},
  pages={1--7},
  year={2013},
  publisher={Wiley Online Library}
}

@article{sharma2024maira,
  title={MAIRA-Seg: Enhancing Radiology Report Generation with Segmentation-Aware Multimodal Large Language Models},
  author={Sharma, Harshita and Salvatelli, Valentina and Srivastav, Shaury and Bouzid, Kenza and Bannur, Shruthi and Castro, Daniel C and Ilse, Maximilian and Bond-Taylor, Sam and Ranjit, Mercy Prasanna and Falck, Fabian and others},
  journal={arXiv preprint arXiv:2411.11362},
  year={2024}
}

@article{zhang2024radgenome,
  title={Radgenome-chest ct: A grounded vision-language dataset for chest ct analysis},
  author={Zhang, Xiaoman and Wu, Chaoyi and Zhao, Ziheng and Lei, Jiayu and Zhang, Ya and Wang, Yanfeng and Xie, Weidi},
  journal={arXiv preprint arXiv:2404.16754},
  year={2024}
}

@article{wu2023towards,
  title={Towards generalist foundation model for radiology by leveraging web-scale 2D\&3D medical data},
  author={Wu, Chaoyi and Zhang, Xiaoman and Zhang, Ya and Wang, Yanfeng and Xie, Weidi},
  journal={arXiv preprint arXiv:2308.02463},
  year={2023}
}

@article{xin2025med3dvlm,
  title={Med3dvlm: An efficient vision-language model for 3d medical image analysis},
  author={Xin, Yu and Ates, Gorkem Can and Gong, Kuang and Shao, Wei},
  journal={IEEE Journal of Biomedical and Health Informatics},
  year={2025},
  publisher={IEEE}
}

@inproceedings{shi2025medm,
  title={Medm-vl: What makes a good medical lvlm?},
  author={Shi, Yiming and Yang, Shaoshuai and Zhu, Xun and Wang, Haoyu and Fu, Xiangling and Li, Miao and Wu, Ji},
  booktitle={International Workshop on Agentic AI for Medicine},
  pages={290--299},
  year={2025},
  organization={Springer}
}

@inproceedings{eslami2023pubmedclip,
  title={Pubmedclip: How much does clip benefit visual question answering in the medical domain?},
  author={Eslami, Sedigheh and Meinel, Christoph and De Melo, Gerard},
  booktitle={Findings of the Association for Computational Linguistics: EACL 2023},
  pages={1181--1193},
  year={2023}
}

@article{perez2025exploring,
  title={Exploring scalable medical image encoders beyond text supervision},
  author={P{\'e}rez-Garc{\'\i}a, Fernando and Sharma, Harshita and Bond-Taylor, Sam and Bouzid, Kenza and Salvatelli, Valentina and Ilse, Maximilian and Bannur, Shruthi and Castro, Daniel C and Schwaighofer, Anton and Lungren, Matthew P and others},
  journal={Nature Machine Intelligence},
  pages={1--12},
  year={2025},
  publisher={Nature Publishing Group UK London}
}

@article{chen2024next,
  title={Next Token Prediction Towards Multimodal Intelligence: A Comprehensive Survey},
  author={Chen, Liang and Wang, Zekun and Ren, Shuhuai and Li, Lei and Zhao, Haozhe and Li, Yunshui and Cai, Zefan and Guo, Hongcheng and Zhang, Lei and Xiong, Yizhe and others},
  journal={arXiv preprint arXiv:2412.18619},
  year={2024}
}

@article{hamamci2024developing,
  title={Developing Generalist Foundation Models from a Multimodal Dataset for 3D Computed Tomography},
  author={Hamamci, Ibrahim Ethem and Er, Sezgin and Almas, Furkan and Simsek, Ayse Gulnihan and Esirgun, Sevval Nil and Dogan, Irem and Dasdelen, Muhammed Furkan and Durugol, Omer Faruk and Wittmann, Bastian and Amiranashvili, Tamaz and others},
  journal={arXiv preprint arXiv:2403.17834},
  year={2024}
}

@article{wasserthal2023totalsegmentator,
  title={TotalSegmentator: robust segmentation of 104 anatomic structures in CT images},
  author={Wasserthal, Jakob and Breit, Hanns-Christian and Meyer, Manfred T and Pradella, Maurice and Hinck, Daniel and Sauter, Alexander W and Heye, Tobias and Boll, Daniel T and Cyriac, Joshy and Yang, Shan and others},
  journal={Radiology: Artificial Intelligence},
  volume={5},
  number={5},
  pages={e230024},
  year={2023},
  publisher={Radiological Society of North America}
}

@article{lee2024read,
  title={Read Like a Radiologist: Efficient Vision-Language Model for 3D Medical Imaging Interpretation},
  author={Lee, Changsun and Park, Sangjoon and Shin, Cheong-Il and Choi, Woo Hee and Park, Hyun Jeong and Lee, Jeong Eun and Ye, Jong Chul},
  journal={arXiv preprint arXiv:2412.13558},
  year={2024}
}

@article{shi2024med,
  title={Med-2E3: A 2D-Enhanced 3D Medical Multimodal Large Language Model},
  author={Shi, Yiming and Zhu, Xun and Hu, Ying and Guo, Chenyi and Li, Miao and Wu, Ji},
  journal={arXiv preprint arXiv:2411.12783},
  year={2024}
}

@article{Huang2024LITALI,
  title={LITA: Language Instructed Temporal-Localization Assistant},
  author={De-An Huang and Shijia Liao and Subhashree Radhakrishnan and Hongxu Yin and Pavlo Molchanov and Zhiding Yu and Jan Kautz},
  journal={ArXiv},
  year={2024},
  volume={abs/2403.19046},
  url={https://api.semanticscholar.org/CorpusID:268733134}
}

@article{Xu2024SlowFastLLaVAAS,
  title={SlowFast-LLaVA: A Strong Training-Free Baseline for Video Large Language Models},
  author={Mingze Xu and Mingfei Gao and Zhe Gan and Hong-You Chen and Zhengfeng Lai and Haiming Gang and Kai Kang and Afshin Dehghan},
  journal={ArXiv},
  year={2024},
  volume={abs/2407.15841},
  url={https://api.semanticscholar.org/CorpusID:271329151}
}

@inproceedings{feichtenhofer2019slowfast,
  title={Slowfast networks for video recognition},
  author={Feichtenhofer, Christoph and Fan, Haoqi and Malik, Jitendra and He, Kaiming},
  booktitle={Proceedings of the IEEE/CVF international conference on computer vision},
  pages={6202--6211},
  year={2019}
}

@article{Zhang2023GPT4RoIIT,
  title={GPT4RoI: Instruction Tuning Large Language Model on Region-of-Interest},
  author={Shilong Zhang and Pei Sun and Shoufa Chen and Min Xiao and Wenqi Shao and Wenwei Zhang and Kai Chen and Ping Luo},
  journal={ArXiv},
  year={2023},
  volume={abs/2307.03601},
  url={https://api.semanticscholar.org/CorpusID:259375716}
}

@article{Ma2024GromaLV,
  title={Groma: Localized Visual Tokenization for Grounding Multimodal Large Language Models},
  author={Chuofan Ma and Yi Jiang and Jiannan Wu and Zehuan Yuan and Xiaojuan Qi},
  journal={ArXiv},
  year={2024},
  volume={abs/2404.13013},
  url={https://api.semanticscholar.org/CorpusID:269283071}
}

@article{Guo2024RegionGPTTR,
  title={RegionGPT: Towards Region Understanding Vision Language Model},
  author={Qiushan Guo and Shalini De Mello and Hongxu Yin and Wonmin Byeon and Ka Chun Cheung and Yizhou Yu and Ping Luo and Sifei Liu},
  journal={2024 IEEE/CVF Conference on Computer Vision and Pattern Recognition (CVPR)},
  year={2024},
  pages={13796-13806},
  url={https://api.semanticscholar.org/CorpusID:268247535}
}

@inproceedings{huang2024refer,
  title={A refer-and-ground multimodal large language model for biomedicine},
  author={Huang, Xiaoshuang and Huang, Haifeng and Shen, Lingdong and Yang, Yehui and Shang, Fangxin and Liu, Junwei and Liu, Jia},
  booktitle={International Conference on Medical Image Computing and Computer-Assisted Intervention},
  pages={399--409},
  year={2024},
  organization={Springer}
}

@misc{zhou2025medversageneralistfoundationmodel,
      title={MedVersa: A Generalist Foundation Model for Medical Image Interpretation}, 
      author={Hong-Yu Zhou and Julián Nicolás Acosta and Subathra Adithan and Suvrankar Datta and Eric J. Topol and Pranav Rajpurkar},
      year={2025},
      eprint={2405.07988},
      archivePrefix={arXiv},
      primaryClass={cs.CV},
      url={https://arxiv.org/abs/2405.07988}, 
}

@misc{wang2025interpretablebilingualmultimodallarge,
      title={Interpretable Bilingual Multimodal Large Language Model for Diverse Biomedical Tasks}, 
      author={Lehan Wang and Haonan Wang and Honglong Yang and Jiaji Mao and Zehong Yang and Jun Shen and Xiaomeng Li},
      year={2025},
      eprint={2410.18387},
      archivePrefix={arXiv},
      primaryClass={cs.CV},
      url={https://arxiv.org/abs/2410.18387}, 
}

@article{Bannur2024MAIRA2GR,
  title={MAIRA-2: Grounded Radiology Report Generation},
  author={Shruthi Bannur and Kenza Bouzid and Daniel C. Castro and Anton Schwaighofer and Sam Bond-Taylor and Maximilian Ilse and Fernando P'erez-Garc'ia and Valentina Salvatelli and Harshita Sharma and Felix Meissen and Mercy Prasanna Ranjit and Shaury Srivastav and Julia Gong and Fabian Falck and Ozan Oktay and Anja Thieme and Matthew P. Lungren and Maria Teodora Wetscherek and Javier Alvarez-Valle and Stephanie L. Hyland},
  journal={ArXiv},
  year={2024},
  volume={abs/2406.04449},
  url={https://api.semanticscholar.org/CorpusID:270357817}
}

@article{Tanida2023InteractiveAE,
  title={Interactive and Explainable Region-guided Radiology Report Generation},
  author={Tim Tanida and Philip M{\"u}ller and Georgios Kaissis and Daniel Rueckert},
  journal={2023 IEEE/CVF Conference on Computer Vision and Pattern Recognition (CVPR)},
  year={2023},
  pages={7433-7442},
  url={https://api.semanticscholar.org/CorpusID:258179419}
}

@article{Yuan2023OspreyPU,
  title={Osprey: Pixel Understanding with Visual Instruction Tuning},
  author={Yuqian Yuan and Wentong Li and Jian Liu and Dongqi Tang and Xinjie Luo and Chi Qin and Lei Zhang and Jianke Zhu},
  journal={2024 IEEE/CVF Conference on Computer Vision and Pattern Recognition (CVPR)},
  year={2023},
  pages={28202-28211},
  url={https://api.semanticscholar.org/CorpusID:266335219}
}

@article{Chen2024LargeLM,
  title={Large Language Model with Region-guided Referring and Grounding for CT Report Generation},
  author={Zhixuan Chen and Yequan Bie and Haibo Jin and Hao Chen},
  journal={ArXiv},
  year={2024},
  volume={abs/2411.15539},
  url={https://api.semanticscholar.org/CorpusID:274234743}
}

@article{liu2023pre,
  title={Pre-train, prompt, and predict: A systematic survey of prompting methods in natural language processing},
  author={Liu, Pengfei and Yuan, Weizhe and Fu, Jinlan and Jiang, Zhengbao and Hayashi, Hiroaki and Neubig, Graham},
  journal={ACM computing surveys},
  volume={55},
  number={9},
  pages={1--35},
  year={2023},
  publisher={ACM New York, NY}
}

@article{Tong2024EyesWS,
  title={Eyes Wide Shut? Exploring the Visual Shortcomings of Multimodal LLMs},
  author={Shengbang Tong and Zhuang Liu and Yuexiang Zhai and Yi Ma and Yann LeCun and Saining Xie},
  journal={2024 IEEE/CVF Conference on Computer Vision and Pattern Recognition (CVPR)},
  year={2024},
  pages={9568-9578},
  url={https://api.semanticscholar.org/CorpusID:266976992}
}

@article{Jin2023PromptMRGDP,
  title={PromptMRG: Diagnosis-Driven Prompts for Medical Report Generation},
  author={Haibo Jin and Haoxuan Che and Yi-Mou Lin and Haoxing Chen},
  journal={ArXiv},
  year={2023},
  volume={abs/2308.12604},
  url={https://api.semanticscholar.org/CorpusID:261100888}
}

@article{Chen2024DiaLLaMATL,
  title={Dia-LLaMA: Towards Large Language Model-driven CT Report Generation},
  author={Zhixuan Chen and Luyang Luo and Yequan Bie and Hao Chen},
  journal={ArXiv},
  year={2024},
  volume={abs/2403.16386},
  url={https://api.semanticscholar.org/CorpusID:268681763}
}

@article{Yeh2024InsightAM,
  title={Insight: A Multi-modal Diagnostic Pipeline Using LLMs for Ocular Surface Disease Diagnosis},
  author={Chun-Hsiao Yeh and Jiayun Wang and Andrew D. Graham and Andrea J. Liu and Bo Tan and Yubei Chen and Yi Ma and Meng C. Lin},
  journal={ArXiv},
  year={2024},
  volume={abs/2410.00292},
  url={https://api.semanticscholar.org/CorpusID:273022716}
}

@article{Bassi2025RadGPTC3,
  title={RadGPT: Constructing 3D Image-Text Tumor Datasets},
  author={Pedro R.A.S. Bassi and Mehmet Can Yavuz and Kang Wang and Xiaoxi Chen and Wenxuan Li and Sergio Decherchi and Andrea Cavalli and Yang Yang and Alan L. Yuille and Zongwei Zhou},
  journal={ArXiv},
  year={2025},
  volume={abs/2501.04678},
  url={https://api.semanticscholar.org/CorpusID:275358308}
}

@inproceedings{Li2023EvaluatingOH,
  title={Evaluating Object Hallucination in Large Vision-Language Models},
  author={Yifan Li and Yifan Du and Kun Zhou and Jinpeng Wang and Wayne Xin Zhao and Ji-rong Wen},
  booktitle={Conference on Empirical Methods in Natural Language Processing},
  year={2023},
  url={https://api.semanticscholar.org/CorpusID:258740697}
}

@article{gao2024anatomy,
  title={Anatomy-Guided Radiology Report Generation with Pathology-Aware Regional Prompts},
  author={Gao, Yijian and Marshall, Dominic and Xing, Xiaodan and Ning, Junzhi and Papanastasiou, Giorgos and Yang, Guang and Komorowski, Matthieu},
  journal={arXiv preprint arXiv:2411.10789},
  year={2024}
}

@article{Claessens2015EarlyCC,
  title={Early Chest Computed Tomography Scan to Assist Diagnosis and Guide Treatment Decision for Suspected Community-acquired Pneumonia.},
  author={Yann Erick Claessens and Marie-Pierre Debray and Florence Tubach and Anne Laure Brun and Blandine Rammaert and Pierre Hausfater and Jean-Marc Naccache and Patrick Ray and Christophe Choquet and Marie France Carette and Charles Mayaud and Catherine Leport and Xavier Duval},
  journal={American journal of respiratory and critical care medicine},
  year={2015},
  volume={192 8},
  pages={
          974-82
        },
  url={https://api.semanticscholar.org/CorpusID:38597236}
}

@article{Marcovici2014JournalCS,
  title={Journal Club: Structured radiology reports are more complete and more effective than unstructured reports.},
  author={Peter Andrew Marcovici and George Albert Taylor},
  journal={AJR. American journal of roentgenology},
  year={2014},
  volume={203 6},
  pages={
          1265-71
        },
  url={https://api.semanticscholar.org/CorpusID:207326910}
}

@article{Cereser2024AssessingTI,
  title={Assessing the impact of structured reporting on learning how to report lung cancer staging CT: A triple cohort study on inexperienced readers.},
  author={Lorenzo Cereser and Francesco Cortiula and C. Simiele and Valeria Peruzzi and Martina Bortolot and Annarita Tullio and Giuseppe Como and Chiara Zuiani and Rossano Girometti},
  journal={European journal of radiology},
  year={2024},
  volume={171},
  pages={
          111291
        },
  url={https://api.semanticscholar.org/CorpusID:266909282}
}

@inproceedings{papineni-etal-2002-bleu,
    title = "{B}leu: a Method for Automatic Evaluation of Machine Translation",
    author = "Papineni, Kishore  and
      Roukos, Salim  and
      Ward, Todd  and
      Zhu, Wei-Jing",
    editor = "Isabelle, Pierre  and
      Charniak, Eugene  and
      Lin, Dekang",
    booktitle = "Proceedings of the 40th Annual Meeting of the Association for Computational Linguistics",
    month = jul,
    year = "2002",
    address = "Philadelphia, Pennsylvania, USA",
    publisher = "Association for Computational Linguistics",
    url = "https://aclanthology.org/P02-1040/",
    doi = "10.3115/1073083.1073135",
    pages = "311--318"
}

@article{macmahon2017guidelines,
  title={Guidelines for management of incidental pulmonary nodules detected on CT images: from the Fleischner Society 2017},
  author={MacMahon, Heber and Naidich, David P and Goo, Jin Mo and Lee, Kyung Soo and Leung, Ann NC and Mayo, John R and Mehta, Atul C and Ohno, Yoshiharu and Powell, Charles A and Prokop, Mathias and others},
  journal={Radiology},
  volume={284},
  number={1},
  pages={228--243},
  year={2017},
  publisher={Radiological Society of North America}
}

@inproceedings{lin-2004-rouge,
    title = "{ROUGE}: A Package for Automatic Evaluation of Summaries",
    author = "Lin, Chin-Yew",
    booktitle = "Text Summarization Branches Out",
    month = jul,
    year = "2004",
    address = "Barcelona, Spain",
    publisher = "Association for Computational Linguistics",
    url = "https://aclanthology.org/W04-1013/",
    pages = "74--81"
}

@article{Ostmeier2024GREENGR,
  title={GREEN: Generative Radiology Report Evaluation and Error Notation},
  author={Sophie Ostmeier and Justin Xu and Zhihong Chen and Maya Varma and Louis Blankemeier and Christian Bluethgen and Arne Edward Michalson and Michael E. Moseley and Curtis P. Langlotz and Akshay S. Chaudhari and Jean-Benoit Delbrouck},
  journal={ArXiv},
  year={2024},
  volume={abs/2405.03595},
  url={https://api.semanticscholar.org/CorpusID:269605082}
}

@inproceedings{Banerjee2005METEORAA,
  title={METEOR: An Automatic Metric for MT Evaluation with Improved Correlation with Human Judgments},
  author={Satanjeev Banerjee and Alon Lavie},
  booktitle={IEEvaluation@ACL},
  year={2005},
  url={https://api.semanticscholar.org/CorpusID:7164502}
}

@article{Hu2021LoRALA,
  title={LoRA: Low-Rank Adaptation of Large Language Models},
  author={J. Edward Hu and Yelong Shen and Phillip Wallis and Zeyuan Allen-Zhu and Yuanzhi Li and Shean Wang and Weizhu Chen},
  journal={ArXiv},
  year={2021},
  volume={abs/2106.09685},
  url={https://api.semanticscholar.org/CorpusID:235458009}
}

@article{Loshchilov2017FixingWD,
  title={Fixing Weight Decay Regularization in Adam},
  author={Ilya Loshchilov and Frank Hutter},
  journal={ArXiv},
  year={2017},
  volume={abs/1711.05101},
  url={https://api.semanticscholar.org/CorpusID:3312944}
}

@article{Rajbhandari2019ZeROMO,
  title={ZeRO: Memory optimizations Toward Training Trillion Parameter Models},
  author={Samyam Rajbhandari and Jeff Rasley and Olatunji Ruwase and Yuxiong He},
  journal={SC20: International Conference for High Performance Computing, Networking, Storage and Analysis},
  year={2019},
  pages={1-16},
  url={https://api.semanticscholar.org/CorpusID:269617042}
}

@article{khattab2023dspy,
  title={Dspy: Compiling declarative language model calls into self-improving pipelines},
  author={Khattab, Omar and Singhvi, Arnav and Maheshwari, Paridhi and Zhang, Zhiyuan and Santhanam, Keshav and Vardhamanan, Sri and Haq, Saiful and Sharma, Ashutosh and Joshi, Thomas T and Moazam, Hanna and others},
  journal={arXiv preprint arXiv:2310.03714},
  year={2023}
}

@article{grattafiori2024llama,
  title={The llama 3 herd of models},
  author={Grattafiori, Aaron and Dubey, Abhimanyu and Jauhri, Abhinav and Pandey, Abhinav and Kadian, Abhishek and Al-Dahle, Ahmad and Letman, Aiesha and Mathur, Akhil and Schelten, Alan and Vaughan, Alex and others},
  journal={arXiv preprint arXiv:2407.21783},
  year={2024}
}

@article{hamamci2025crg,
  title={CRG Score: A Distribution-Aware Clinical Metric for Radiology Report Generation},
  author={Hamamci, Ibrahim Ethem and Er, Sezgin and Shit, Suprosanna and Reynaud, Hadrien and Kainz, Bernhard and Menze, Bjoern},
  journal={arXiv preprint arXiv:2505.17167},
  year={2025}
}

@article{sellergren2025medgemma,
  title={Medgemma technical report},
  author={Sellergren, Andrew and Kazemzadeh, Sahar and Jaroensri, Tiam and Kiraly, Atilla and Traverse, Madeleine and Kohlberger, Timo and Xu, Shawn and Jamil, Fayaz and Hughes, C{\'\i}an and Lau, Charles and others},
  journal={arXiv preprint arXiv:2507.05201},
  year={2025}
}

\clearpage
\setcounter{page}{1}
\setcounter{section}{0}
\setcounter{figure}{0}
\setcounter{table}{0}

\begin{center}
    {\Large \bfseries Supplementary Material\par}
\end{center}

\renewcommand{\thefigure}{S\arabic{figure}}
\renewcommand{\thetable}{S\arabic{table}}

\section{Implementation Details} 
\label{supplementary_sec1}
We preprocess 3D CT volumes by resampling to a uniform voxel spacing of $1 \times 1 \times 3$ $mm$. To standardize the input, we perform foreground cropping followed by center spatial cropping and padding to a fixed resolution of $448 \times 448 \times 128$. To ensure compatibility with the pre-trained MedSigLIP vision encoder \cite{sellergren2025medgemma}, we adopt its specific 3-channel multi-windowing strategy for intensity normalization. Specifically, the Hounsfield Unit (HU) ranges for bone-lung, soft tissue, and brain are mapped into an RGB-like representation, with each channel normalized to $[-1, 1]$. For the multimodal connector, we use a linear layer. LLaMA-3.1-8B \cite{grattafiori2024llama} serves as the language model, with a maximum sequence length of 8,192. For parameter-efficient fine-tuning, we apply LoRA \cite{Hu2021LoRALA} with rank $r = 128$, $\alpha = 256$, and a dropout ratio of 0.1. MLLM training consists of two stages. In the first stage, we freeze the vision encoder and LLM while fine-tuning the multimodal connector and mask projector on an image-report dataset, using an effective batch size of 48, a learning rate of $10^{-3}$, and 6 epochs with warmup and cosine decay. In the second stage, we fine-tune the multimodal connector, the mask projector, and the LLM using a structured region-specific report dataset, with an effective batch size of 48, a learning rate of $2 \times 10^{-5}$, and 6 epochs, again with warmup and cosine decay. All models are trained using the AdamW optimizer \cite{Loshchilov2017FixingWD} with ZeRO stage 3 \cite{Rajbhandari2019ZeROMO}. Our implementation is based on PyTorch, and training is conducted in parallel on NVIDIA A100 GPUs, each with 80 GB of memory. Bootstrap resampling (10,000 replicates) provided 95\% confidence intervals.

\section{Extraction of Lesion Attributes}
\label{supplementary_sec2}
As shown in Supplementary Algorithm~\ref{supple_alg1}, we extract localized quantitative attributes from 3D segmentation masks to provide structured clinical evidence. The input masks are first normalized to a uniform spacing of $1.0 \times 1.0 \times 1.0$ $mm$. To assign pathological findings to specific anatomical regions, we perform a voxel-wise intersection between lesion masks ($M_{les}$) and organ subregion masks ($M_{loc}$), such as specific lung lobes or kidney sides. To ensure robustness against segmentation artifacts, we apply a Connected Component Analysis (CCA) and retain only the largest component ($M_{single}$) for each ROI. We then derive two primary metrics: (i) Volume, calculated by the product of the voxel count and the unit voxel volume; and (ii) Maximum Diameter, determined by the maximum axial length of the 3D bounding box enclosing the lesion. These extracted attributes—volume (mL), diameter (mm), and anatomical location—serve as the quantitative foundation for the subsequent structured report generation.

\section{Region-level Performance}
\label{supplementary_sec3}
As shown in Supplementary Tab.~\ref{supplementary_tab1}, MedRegion-CT consistently outperforms state-of-the-art baselines across almost all major anatomical regions. By selectively aggregating visual features within predefined anatomical masks, the model effectively captures region-specific pathological findings. The only exception is the large airway, where our model achieves highly competitive results but does not yield the highest absolute score. This isolated limitation suggests that while our volumetric sampling strategy excels for dense organs, the combination of a tubular structure and a relatively small volumetric footprint makes the large airway uniquely vulnerable to feature dilution.

\section{Key Organ and Sub-region Taxonomy}
\label{supplementary_sec4}
We define a hierarchical anatomical taxonomy focused on six clinically significant organ systems. Each primary organ is decomposed into fine-grained sub-regions to facilitate localized pathological analysis. Detailed specifications of this anatomical hierarchy are summarized in Supplementary Tab.~\ref{supplementary_tab2}. The current organ selection is optimized for the label space of the TotalSegmentator \cite{wasserthal2023totalsegmentator}, which serves as our backbone for universal segmentation. While this taxonomy aligns with the architectural constraints of the chosen segmentor, our framework is designed with a model-agnostic architecture. This modularity ensures that the proposed pipeline can seamlessly integrate evolving segmentation backbones or be extended to broader anatomical structures as segmentation technologies advance.

\section{Performance Analysis by Lesion Size}
\label{supplementary_sec5}
To thoroughly understand the detection capabilities of our framework, we stratified lung nodule detection performance by lesion size using the internal RadGenome-Chest CT dataset, as detailed in Supplementary Tab.~\ref{supplementary_tab3}. MedRegion-CT achieves the highest detection rates for mid-to-large nodules compared to existing baselines. However, a noticeable performance drop occurs for very small nodules ($\le 5\text{ mm}^3$). Our case-level analysis reveals that this limitation primarily stems from false negatives generated by the universal segmentation model (TotalSegmentator), which failed to identify the target lesion in 110 out of 160 micro-nodule cases. Thus, the model's sensitivity lower-bound is inherently constrained by the detection limits of the upstream segmentor.

\section{Radiologist Reader Study}
\label{supplementary_sec6}
To further validate the clinical efficacy of our model in a real-world external setting, we conducted a blinded radiologist reader study on the AMC dataset. As summarized in Supplementary Tab.~\ref{supplementary_tab4}, a board-eligible senior radiology resident evaluated the generated reports for 50 randomly sampled cases. The generated reports were subjectively scored on a scale of 1 to 10 against the paired ground-truth reports, with a strict focus on the accurate capture of key clinical findings and the absence of hallucinations. MedRegion-CT substantially outperforms all baseline models, achieving an average score of 6.98 ($\pm$1.41), compared to the second-best score of 4.44 ($\pm$2.01) by Med3DVLM. This substantial margin in human expert evaluation corroborates our LM-based quantitative metrics, demonstrating that our framework produces highly reliable and clinically accurate reports even under cross-institutional domain shifts, where surface-level lexical metrics often fail to reflect true diagnostic fidelity.

\section{Robustness to Segmentation Noise}
\label{supplementary_sec7}
To evaluate the dependency of MedRegion-CT on the quality of pseudo-masks, we conducted a systematic segmentation noise stress test (Supplementary Tab.~\ref{supplementary_tab5}). As our framework relies on both \textit{organ} and \textit{lesion} masks to guide visual representation learning, performance degradation under mask distortion is expected. Interestingly, however, we observe that distorting the organ masks results in a substantially larger performance drop compared to distorting the lesion masks. We attribute this discrepancy to the scope of each mask's influence within our architecture. Specifically, the organ masks provide critical ROI guidance for two foundational modules: both the Region-based SlowFast Tokenizer and the Mask-Driven Visual Extractor heavily depend on them to extract and align regional visual features. In contrast, the lesion masks are exclusively utilized by the Lesion Attribute Extractor to derive quantitative textual prompts.

\clearpage

\begin{algorithm}[tb]
\caption{Lesion Attribute Extractor Algorithm}
\label{supple_alg1}
\textbf{Input}: Segmentation masks for anatomical guidance regions ($\mathcal{M}_{loc}$), lesions ($\mathcal{M}_{les}$). \\
\textbf{Output}: Textual prompt list from medical attributes.
\begin{algorithmic}[1]
    \STATE \textbf{function} GET\_MAX\_DIAMETER($mask, s_x, s_y, s_z$)
    \STATE \quad \textbf{if} $mask$ is \text{None} \textbf{or} $\text{sum}(mask) == 0$ \textbf{return} 0
    \STATE \quad $L_x, L_y, L_z \gets \text{lengths of bounding box of } mask \text{ along } x, y, z \text{ axes}$
    \STATE \quad \textbf{return} $\max(L_x \times s_x, L_y \times s_y, L_z \times s_z)$
    \STATE \textbf{end function}

    \STATE $\Phi(\cdot) \gets \text{Operator extracting the largest connected component}$
    \STATE $attr\_prompt\_list \gets [\,]$
    \STATE
    
    \FOR{each target lesion class ($c_{les}$) \textbf{and} mask ($\mathcal{M}_{les}$) \textbf{in} $\{\text{Nodule}, \text{Cyst}, \text{Effusion}\}$}
        \FOR{each anatomical guidance mask ($\mathcal{M}_{loc}$) \textbf{in} $\{\text{Left Upper Lobe}, \dots\}$}
            \STATE $\mathcal{M}_{target} \gets \Phi(\mathcal{M}_{loc} \odot \mathcal{M}_{les})$ \COMMENT{Intersection and extraction}
            \STATE $V \gets \text{sum}(\mathcal{M}_{target}) \times (s_x \times s_y \times s_z)/10^3$
            \STATE $d_{max} \gets \text{GET\_MAX\_DIAMETER}(\mathcal{M}_{target}, s_x, s_y, s_z)$
            \STATE \textbf{Save} $\{V, d_{max}\}$ for the current $\mathcal{M}_{loc}$

            \IF{$V > 0$}
                \IF{$d_{max}$ is not None}
                    \STATE $prompt \gets$ ``A $c_{les}$ with a diameter of $d_{max}$ mm and a volume of $V$ mL is identified in the $\mathcal{M}_{loc}$.''
                \ELSE              
                    \STATE $prompt \gets \text{``A } c_{les} \text{ with a volume of } V \text{ mL}$ \text{ is identified in the }$ \mathcal{M}_{loc}\text{.''}$
                \ENDIF
            \ELSE
                \STATE $prompt \gets \text{``No } c_{les} \text{ is present in the } \mathcal{M}_{loc}\text{.''}$
            \ENDIF
            \STATE $attr\_prompt\_list\text{.append}(prompt)$
        \ENDFOR
    \ENDFOR

    \STATE 
    \RETURN{} $attr\_prompt\_list$
\end{algorithmic}
\end{algorithm}

\begin{table}[tb]
    \caption{Region-wise performance evaluation on a RadGenome-Chest CT structured report generation benchmark. Note: Mediast.: Mediastinum, Musculo.: Musculoskeletal. Best results are highlighted in bold.}    
    \centering
    \resizebox{\textwidth}{!}{
    \begin{tabular}{l cccccc cccccc}
        \toprule
        \multirow{2}{*}{\textbf{Region}} & \multicolumn{6}{c}{\textbf{METEOR}} & \multicolumn{6}{c}{\textbf{GREEN}} \\
        \cmidrule(lr){2-7} \cmidrule(lr){8-13}
        & RadFM & M3D & MedM-VL & Med3DVLM & CT-CHAT & \textbf{Ours} & RadFM & M3D & MedM-VL & Med3DVLM & CT-CHAT & \textbf{Ours} \\
        \midrule
        Lung      & 0.3141 & 0.3316 & 0.3002 & 0.3500 & 0.3067 & \textbf{0.3591} & 0.1805 & 0.2530 & 0.2233 & 0.2840 & 0.2361 & \textbf{0.3098} \\
        Airway    & 0.5100 & 0.5602 & 0.5481 & 0.5528 & \textbf{0.5632} & 0.5444 & 0.6779 & \textbf{0.7460} & 0.7403 & 0.7298 & 0.7446 & 0.7306 \\
        Mediast.  & 0.4212 & 0.4445 & 0.4359 & 0.4513 & 0.4363 & \textbf{0.4800} & 0.5119 & 0.5873 & 0.5847 & 0.5877 & 0.5921 & \textbf{0.6214} \\
        Heart     & 0.3780 & 0.4191 & 0.4020 & 0.4178 & 0.4025 & \textbf{0.4633} & 0.4521 & 0.5052 & 0.4928 & 0.4986 & 0.4971 & \textbf{0.5512} \\
        Musculo.  & 0.4054 & 0.4512 & 0.4560 & 0.4489 & 0.4534 & \textbf{0.4646} & 0.3458 & 0.4182 & 0.4008 & 0.4176 & 0.4306 & \textbf{0.4340} \\
        Abdomen   & 0.3604 & 0.3920 & 0.3928 & 0.4086 & 0.3951 & \textbf{0.4587} & 0.3043 & 0.3863 & 0.3927 & 0.4133 & 0.3910 & \textbf{0.4823} \\
        \bottomrule
    \end{tabular}%
    }
    \label{supplementary_tab1}
\end{table}

\begin{table}[tb]
\captionsetup{justification=raggedright,singlelinecheck=false}
\caption{List of major organs and their corresponding sub-region structures.}
\label{supplementary_tab2}
\centering
{\small
\begin{tabularx}{\linewidth}{@{}c >{\raggedright\arraybackslash}X@{}}
\toprule
\textbf{No.} & \textbf{Category and Sub-Region Structures} \\
\midrule
1 & \textbf{Lung Parenchyma:} Left lung upper lobe; Left lung lower lobe; Right lung upper lobe; Right lung middle lobe; Right lung lower lobe \\
2 & \textbf{Large Airway:} Trachea \\
3 & \textbf{Mediastinum:} Esophagus; Brachiocephalic trunk; Right subclavian artery; Left subclavian artery; Left brachiocephalic vein; Right brachiocephalic vein \\
4 & \textbf{Heart and Great Vessels:} Heart; Aorta; Pulmonary vein; Left atrial appendage; Superior vena cava; Inferior vena cava \\
5 & \textbf{Abdominal Organs:} Spleen; Right kidney; Left kidney; Gallbladder; Liver; Stomach; Pancreas; Right adrenal gland; Left adrenal gland; Small bowel; Duodenum \\
6 & \textbf{Musculoskeletal Structures:} Cervical vertebrae C6--C7; Thoracic vertebrae T1--T12; Lumbar vertebrae L1--L4; Spinal cord; Left humerus; Right humerus; Left scapula; Right scapula; Left clavicle; Right clavicle; Left ribs 1--12; Right ribs 1--12; Sternum; Costal cartilages \\
\bottomrule
\end{tabularx}
}
\end{table}

\begin{table}[tb]
    \caption{\footnotesize Lung nodule detection performance across different size categories. We report the detection rate among all nodule patients. Analysis is based on TotalSegmentator and RadBERT predictions. Best results are highlighted in bold.}
    \centering
    \resizebox{\textwidth}{!}{
    \begin{tabular}{l ccc ccc}
        \toprule
        \textbf{Category} & RadFM & M3D & CT-CHAT & MedM-VL & Med3DVLM & Ours \\
        \midrule
        \textbf{Small}      & 0.3492 & 0.3016 & \textbf{0.3651} & 0.2937 & 0.2857 & 0.1270 \\
        \textbf{Lower-Mid}  & 0.2952 & 0.3065 & 0.3065 & 0.3065 & 0.2903 & \textbf{0.3871} \\
        \textbf{Upper-Mid}  & 0.3307 & 0.4016 & 0.3150 & 0.3228 & 0.4094 & \textbf{0.4882} \\
        \textbf{Large}      & 0.3175 & 0.4206 & 0.4127 & 0.3730 & \textbf{0.4286} & \textbf{0.4286} \\
        \bottomrule
    \end{tabular}%
    }
    \label{supplementary_tab3}
\end{table}


\begin{table}[tb]
    \caption{\footnotesize Radiologist Reader Study results. A board-eligible senior radiology resident audited 50 randomly sampled cases under blinding. Findings were evaluated on a 1–10 scale, scored against the paired ground-truth reports to account for key findings and hallucinations. Best results are highlighted in bold.}
    \centering
    \label{supplementary_tab4}
    \resizebox{\textwidth}{!}{
    \begin{tabular}{l cccccc}
        \toprule
        \textbf{Reader Study} & RadFM & M3D & MedM-VL & Med3DVLM & CT-CHAT & Ours \\
        \midrule
        \textbf{Score {\scriptsize (Avg. $\pm$ Std.)}} & 3.1875 & 3.7292 & 3.0625 & 4.4375 & 2.9792 & \textbf{6.9792} \\
        & {\scriptsize $\pm$1.5389} & {\scriptsize $\pm$1.6469} & {\scriptsize $\pm$1.8266} & {\scriptsize $\pm$2.0096} & {\scriptsize $\pm$1.5366} & {\scriptsize $\pm$1.4065} \\
        \bottomrule
    \end{tabular}%
    } 
\end{table}


\begin{table}[tb]
    \caption{\footnotesize Ablation study on segmentation noise robustness. We evaluate the impact of mask quality by comparing performance with accurate (\textcolor{ForestGreen}{Good}) and perturbed (\textcolor{Red}{Distorted}) masks for both lesions and organs.}
    \centering
    \resizebox{0.6\textwidth}{!}{%
    \begin{tabular}{l c cc}
        \toprule
        \textbf{Lesion Mask} & \textbf{Organ Mask} & \textbf{BLEU} & \textbf{CRG} \\
        \midrule
        \textcolor{ForestGreen}{Good}      & \textcolor{ForestGreen}{Good}      & \textbf{0.3273} & \textbf{0.3861} \\
        \textcolor{ForestGreen}{Good}      & \textcolor{Red}{Distorted} & 0.2252          & 0.3654 \\
        \textcolor{Red}{Distorted} & \textcolor{ForestGreen}{Good}      & 0.2986          & 0.3799 \\
        \textcolor{Red}{Distorted} & \textcolor{Red}{Distorted} & 0.2364          & 0.3604 \\
        \bottomrule
    \end{tabular}%
    }
    \label{supplementary_tab5}
\end{table}

\begin{figure}[tb]
\includegraphics[width=\textwidth, trim=0 16 0 0, clip]{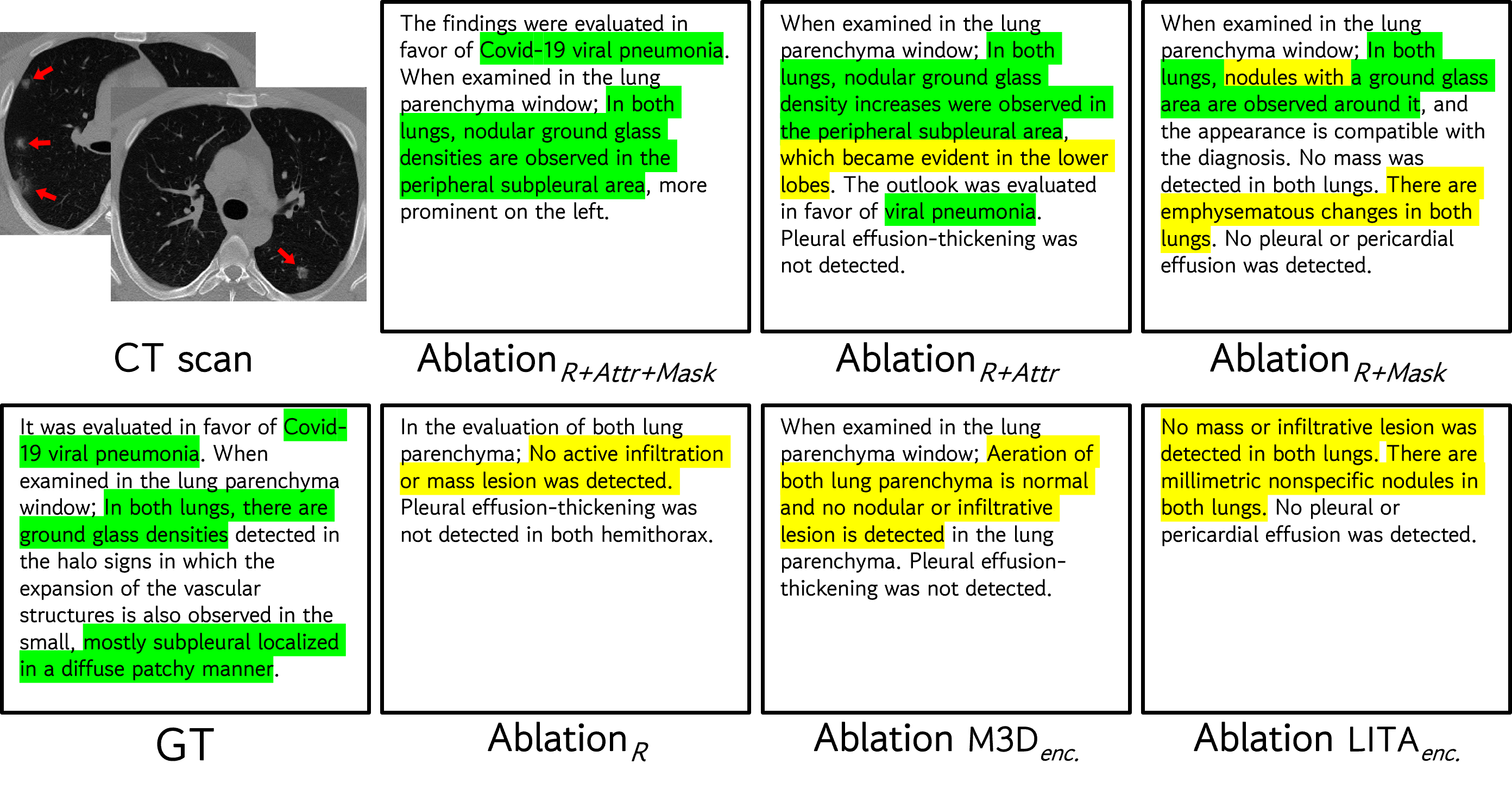}
\caption{Qualitative ablation analysis of MedRegion-CT components on a case of COVID-19 viral pneumonia. Text is color-coded to indicate findings consistent with the ground truth (\textcolor{green}{green}) 
and findings that are inconsistent with the ground truth or represent hallucinations (\textcolor{yellow}{yellow}). For clarity, only the lung section of the report is visualized. The full model most accurately identifies the subpleural localized patterns.}
\label{supple_fig1}
\end{figure}

\end{document}